\documentclass[12pt]{iopart}

\usepackage{graphics}
\usepackage{graphicx}
\usepackage{graphics}
\usepackage{iopams}       

\usepackage{upgreek}     
\usepackage{setstack} 
\usepackage{upgreek}
\usepackage{color}
\usepackage[colorlinks,citecolor=blue,urlcolor=blue,linkcolor=blue]{hyperref}

\usepackage{units}
\usepackage{multirow}
\usepackage{booktabs} 

\newcommand{\bb}[1]{$\rm #1\nu\beta\beta$}
\newcommand{\gadash}[1]{$\gamma$ #1}

\begin{document}

%
%
%
\newcommand{\versionabb}  {version 10.6 dated 20141219}
\newcommand{\ctsper}      {cts/(keV$\cdot$kg$\cdot$yr)}
\newcommand{\ctsperee}    {cts/(keV$_{ee}\cdot$kg$\cdot$yr)}
\newcommand{\ctsperrec}   {cts/(keV$_{rec}\cdot$kg$\cdot$yr)}
\newcommand{\zctsper}     {{$10^{-2}$~cts/(keV$\cdot$kg$\cdot$yr)}}
\newcommand{\tctsper}     {{$10^{-2}$~cts/(keV$\cdot$kg$\cdot$yr)}}
\newcommand{\pIbi}        {{$10^{-2}$~cts/(keV$\cdot$kg$\cdot$yr)}}
\newcommand{\dctsper}     {{$10^{-3}$~cts/(keV$\cdot$kg$\cdot$yr)}}
\newcommand{\pIIbi}       {{$10^{-3}$~cts/(keV$\cdot$kg$\cdot$yr)}}
\newcommand{\biperton}    {{1~cts/(keV$\cdot$t$\cdot$yr)}}
\newcommand{\vctsper}     {{$10^{-4}$~cts/(keV$\cdot$kg$\cdot$yr)}}
\newcommand{\ctsperx}     {$\frac{10^{-3}\rm cts}{\rm keV\cdot kg \cdot yr}$}
\newcommand{\kgy}         {{kg$\cdot$yr}}
\newcommand{\kgyr}        {{kg$\cdot$yr}}
\newcommand{\kevkgyr}     {{keV$\cdot$kg$\cdot$yr}}
\newcommand{\cum}         {{m$^3$}}
\newcommand{\mubq}        {{$\upmu$Bq}}
\newcommand{\mum}         {{$\upmu$m}}
\newcommand{\mus}         {{$\upmu$s}}
\newcommand{\mubqperkg}   {{${\upmu\mathrm{Bq}}/{\mathrm{kg}}$}}
\def\cpowten#1#2{{$#1\cdot10^{#2}$}}
\def\powten#1{{$10^{#1}$}}
\newcommand{\baseT}[2]{\mbox{$#1{\cdot}10^{#2}$}}
\newcommand{\baseTsolo}[1]{$10^{#1}$}
\newcommand{\C}           {$^\circ$C}
\newcommand{\al}          {$\alpha$}
\newcommand{\be}          {$\beta$}
\newcommand{\ga}          {$\gamma$}
\newcommand{\gam}         {$\gamma$}
\newcommand{\gammas}      {$\gamma${s}}
\newcommand{\qbb}         {{$Q_{\beta\beta}$}}
\newcommand{\upqbb}       {Q$_{\upbeta\upbeta}$}
\newcommand{\Qbb}         {{$\text{Q}_{\beta\beta}$}}
\newcommand{\thalfzero}   {${T^{0\nu}_{1/2}}$}
\newcommand{\thalftwo}    {${T^{2\nu}_{1/2}}$}
\newcommand{\thalfmajo}   {{${T^{0\nu \chi }_{1/2}}$}}
\newcommand{\nmez}        {${\cal M}^{0\nu}$}
\newcommand{\nmet}        {${\cal M}^{2\nu}$}
\newcommand{\mbb}         {${\langle m_{\beta\beta}\rangle}$}
\newcommand{\bbno}        {{$0\nu\beta\beta$}}
\newcommand{\onbb}        {{$0\nu\beta\beta$}}
\newcommand{\onbbchi}     {{$0\nu\beta\beta\chi$}}
\newcommand{\nnbb}        {{$2\nu\beta\beta$}}
\newcommand{\uponbb}      {{$0\upnu\upbeta\upbeta$}}
\newcommand{\upnnbb}      {{$2\upnu\upbeta\upbeta$}}
\newcommand{\twonu}       {{$2\nu\beta\beta$}}
\newcommand{\tlive}       {\mbox{$t_{live}$}}
\newcommand{\flive}       {\mbox{$f_{live}$}}
\newcommand{\fge}         {\mbox{$f_{Ge}$}}
\newcommand{\fgesix}      {\mbox{$f_{76}$}}
\newcommand{\fgesixi}     {\mbox{$f_{76,i}$}}
\newcommand{\actmass}     {\mbox{$M_{act}$}}
\newcommand{\ssmass}      {\mbox{$M_{76}$}}
\newcommand{\factmass}    {\mbox{$f_{av}$}}
\newcommand{\factmassi}    {\mbox{$f_{av,i}$}}
\newcommand{\factvol}     {\mbox{$f_{av}$}}
\newcommand{\factvoli}     {\mbox{$f_{av,i}$}}
\newcommand{\subssix}      {\mbox{$_{76}$}}
\newcommand{\subssixi}      {\mbox{$_{76,i}$}}
\newcommand{\subav}       {\mbox{$_{av}$}}
\newcommand{\subavi}       {\mbox{$_{av,i}$}}
\newcommand{\bal}         {\rule{10mm}{0.2mm}}
\newcommand{\balf}        {\rule{10mm}{2mm}}
\newcommand{\tabta}       {\rule[-1,5mm]{0mm}{7,5mm}}
\newcommand{\tabtag}      {\rule[-3mm]{0mm}{9mm}}
\newcommand{\up}          {\rule{0mm}{5mm}}
\newcommand{\down}        {\rule[-2mm]{0mm}{3mm}}
\renewcommand{\etal}        {\textit{et al.}}
\newcommand{\bulitem}     {\item[{\large$\bullet$}]}
\newcommand{\prep}        {\textit{Preprint}}
\newcommand{\PRB}         {Phys.~Rev.~B}
\newcommand{\mpik}        {\mbox{MPI-K}}
\newcommand{\mpip}        {\mbox{MPI-P}}
\newcommand{\gerda}       {\textsc{Gerda}}
\newcommand{\G}           {{\mbox{\textsc{Gerda}}}}
\newcommand{\GERDA}       {\mbox{\textsc{Gerda}}}  
\newcommand{\lngs}        {{\mbox{\textsc{Lngs}}}}
\newcommand{\LNGS}        {{\mbox{\textsc{Lngs}}}}
\newcommand{\WT}          {water tank}
\newcommand{\lar}         {LAr}
\newcommand{\geni}        {{\mbox{\textsc{Genius}}}}
\newcommand{\gdl}         {\textsc{Gdl}}
\newcommand{\gerdalarge}  {\textsc{Gerda-LArGe}}
\newcommand{\bege}        {{\sc BEGe}}
\newcommand{\phaseone}    {Phase~I}
\newcommand{\phasetwo}    {Phase~II}
\newcommand{\phasetwop}   {Phase~II$^+$}

\newcommand{\LArGe}       {\textsc{LArGe}}
\newcommand{\SUB}         {{\mbox{\text{SUB}}}}
\newcommand{\Gerdella}    {{\mbox{\textsc{Gerdella}}}}
\newcommand{\GeMPI}       {Ge\textsc{MPI}}
\newcommand{\GEMPI}       {Ge\textsc{mpi}}
\newcommand{\majorana}    {\textsc{Majorana}}
\newcommand{\Majorana}    {{\mbox{\textsc{Majorana}}}}
\newcommand{\MAJORANA}    {{\sc Majorana}}
\newcommand{\eureca}      {{\sc Eureca}}
\newcommand{\igex}        {\textsc{Igex}}
\newcommand{\IGEX}        {{\mbox{\textsc{Igex}}}}
\newcommand{\hdm}         {\textsc{HdM}}
\newcommand{\HdM}         {\mbox{\textsc{HdM}}}  
\newcommand{\HDM}         {\mbox{\textsc{HdM}}}  
\newcommand{\BX}          {\mbox{\textsc{Borexino}}}
\newcommand{\borex}       {\mbox{\textsc{Borexino}}}
\newcommand{\borexino}       {\mbox{\textsc{Borexino}}}
\newcommand{\EUROBALL}    {\textsc{Euroball}}
\newcommand{\ICARUS}      {\textsc{Icarus}}
\newcommand{\LENS}        {\textsc{Lens}}
\newcommand{\STRAW}       {\textsc{Straw}}
\newcommand{\Gallex}      {\textsc{Gallex}}
\newcommand{\LVD}         {{\mbox{\textsc{Lvd}}}}
\newcommand{\GNO}         {{\mbox{\textsc{Gno}}}}
\newcommand{\KAMLAND}     {{\mbox{\textsc{KAMland}}}}
\newcommand{\CUORI}       {{\mbox{\textsc{Cuoricino}}}}
\newcommand{\CUORE}       {{\mbox{\textsc{Cuore}}}}
\newcommand{\cuore}       {{\mbox{\textsc{Cuore}}}}
\newcommand{\AGATA}       {{\mbox{\textsc{Agata}}}}
\newcommand{\WMAP}        {{\mbox{\textsc{Wmap}}}}
\newcommand{\NEMO}        {{\mbox{\textsc{Nemo}}}}
\newcommand{\KATRIN}      {{\mbox{\textsc{Katrin}}}}
\newcommand{\MOON}        {{\mbox{\textsc{Moon}}}}
\newcommand{\EXO}         {{\mbox{\textsc{Exo}}}}
\newcommand{\GEM}         {{\mbox{\textsc{Gem}}}}
\newcommand{\CAMEO}       {{\mbox{\textsc{Cameo}}}}
\newcommand{\COBRA}       {{\mbox{\textsc{Cobra}}}}
\newcommand{\CRYOGENM}    {{\mbox{\textsc{Cryogenmash}}}}
\newcommand{\xenon}       {{\mbox{\textsc{Xenon}}}}
\newcommand{\Xenon}       {{\mbox{\textsc{Xenon}}}}
\newcommand{\taiga}       {{\mbox{\textsc{Taiga}}}}
\newcommand{\geant}       {\textsc{Geant4}}
\newcommand{\GEANT}       {\textsc{\mbox{{Geant}}}}
\newcommand{\rootv}       {\textsc{Root}}
\newcommand{\CERN}        {{\mbox{\textsc{Cern}}}}
\newcommand{\mage}        {\textsc{MaGe}}
\newcommand{\gelatio}     {\textsc{Gelatio}}
\newcommand{\mgdo}        {\mbox{MGDO}}
\newcommand{\tier}        {\textsc{Tier}}
\newcommand{\gesix}       {{$^{76}$Ge}}
\newcommand{\geseven}     {{$^{77}$Ge}}
\newcommand{\gefour}      {{$^{74}$Ge}}
\newcommand{\gethree}     {{$^{73}$Ge}}
\newcommand{\gess}        {{$^{76}$Ge}}
\newcommand{\gesf}        {{$^{74}$Ge}}
\newcommand{\geenr}       {{$^{\rm enr}$Ge}}          
\newcommand{\genat}       {{$^{\rm nat}$Ge}}
\newcommand{\gedep}       {{$^{\rm dep}$Ge}}
\newcommand{\geox}        {{GeO$_2$}}
\newcommand{\enrgecoax}   {{$^{\rm enr}$Ge-coax}} 
\newcommand{\nuc}[2]      {{$^{#2}$\rm #1}}

\newcommand{\cosix}       {{$^{60}$Co}}
\newcommand{\radzzs}      {{$^{226}$Ra}}
\newcommand{\thzza}       {{$^{228}$Th}}
\newcommand{\tlzna}       {{$^{208}$Tl}}
\newcommand{\kvn}         {{$^{40}$K}}
\newcommand{\kvz}         {{$^{42}$K}}
\newcommand{\Am}          {$^{241}$Am}
\newcommand{\Rn}          {$^{222}$Rn}
\newcommand{\Ra}          {$^{226}$Ra}
\newcommand{\Po}          {$^{210}$Po}
\newcommand{\Ar}          {$^{39}$Ar}
\newcommand{\Kr}          {$^{85}$Kr}
\newcommand{\Xe}          {$^{133}$Xe}
\newcommand{\Ba}          {$^{133}$Ba}
\newcommand{\Bi}          {$^{214}$Bi}
\newcommand{\Yb}          {$^{176}$Yb}
\newcommand{\Fe}          {$^{55}$Fe}
\newcommand{\Th}          {$^{228}$Th}
\newcommand{\Tl}          {$^{208}$Tl}
\newcommand{\Ul}          {$^{235}$U}
\newcommand{\Uh}          {$^{238}$U}
\newcommand{\Be}          {$^7$Be}
\newcommand{\YO}          {Yb$_2$O$_3$}
\newcommand{\Co}          {$^{60}$Co}
\newcommand{\exposure}    {\mbox{$\cal E$}}
\newcommand{\effqual}     {\mbox{$\varepsilon_{qcut}$}}
\newcommand{\effpsd}      {\mbox{$\varepsilon_{psd}$}}
\newcommand{\effres}      {\mbox{$\varepsilon_{res}$}}
\newcommand{\efffep}      {\mbox{$\varepsilon_{fep}$}}
\newcommand{\effquali}     {\mbox{$\varepsilon_{qcut,i}$}}
\newcommand{\effpsdi}      {\mbox{$\varepsilon_{psd,i}$}}
\newcommand{\effresi}      {\mbox{$\varepsilon_{res,i}$}}
\newcommand{\efffepi}      {\mbox{$\varepsilon_{fep,i}$}}
\newcommand{\effmuon}     {\mbox{$\varepsilon_\mu$}}
\newcommand{\effmurej}    {\mbox{$\varepsilon_{rej}$}}
\newcommand{\effmureji}    {\mbox{$\varepsilon_{rej,i}$}}


\title{$2\nu\beta\beta$ decay of $^{76}$Ge into excited states with
                  \protect{\sc Gerda}  Phase~I }

\author{
\protect{\sc Gerda} Collaboration
         \footnote{LNGS, Assergi, Italy; correspondence: gerda-eb@mpi-hd.mpg.de}
~\\[5mm]
  M.~Agostini$^{15}$,
  M.~Allardt$^{4}$,
  A.M.~Bakalyarov$^{13}$,
  M.~Balata$^{1}$,
  I.~Barabanov$^{11}$,
  N.~Barros$^{4}$\footnote{present address: Dept. of Physics and Astronomy,
  Univ. of Pennsylvania, Philadelphia, Pennsylvania, USA},
  L.~Baudis$^{19}$,
  C.~Bauer$^{7}$,
  N.~Becerici-Schmidt$^{14}$,
  E.~Bellotti$^{8,9}$
% ...
}
\address{
$^{1}$\ INFN Laboratori Nazionali del Gran Sasso and Gran Sasso Science Institute, Assergi, Italy\label{ALNGS} \\
$^{2}$\ INFN Laboratori Nazionali del Sud, Catania, Italy\label{CAT} \\
$^{3}$\ Institute of Physics, Jagiellonian University, Cracow, Poland\label{CR} \\
$^{4}$\ Institut f{\"u}r Kern- und Teilchenphysik, Technische Universit{\"a}t Dresden, Dresden, Germany\label{DD} \\
$^{5}$\ Joint Institute for Nuclear Research, Dubna, Russia\label{JINR} \\
$^{6}$\ Institute for Reference Materials and Measurements, Geel, Belgium\label{GEEL} \\
$^{7}$\ Max-Planck-Institut f{\"u}r Kernphysik, Heidelberg, Germany\label{HD} \\
$^{8}$\ Dipartimento di Fisica, Universit{\`a} Milano Bicocca, Milano, Italy\label{MIBF} \\
$^{9}$\ INFN Milano Bicocca, Milano, Italy\label{MIBINFN} \\
$^{10}$\ Dipartimento di Fisica, Universit{\`a} degli Studi di Milano e INFN Milano, Milano, Italy\label{MILUINFN} \\
$^{11}$\ Institute for Nuclear Research of the Russian Academy of Sciences, Moscow, Russia\label{INR} \\
$^{12}$\ Institute for Theoretical and Experimental Physics, Moscow, Russia\label{ITEP} \\
$^{13}$\ National Research Centre ``Kurchatov Institute'', Moscow, Russia\label{KU} \\
$^{14}$\ Max-Planck-Institut f{\"ur} Physik, M{\"u}nchen, Germany\label{MPIP} \\
$^{15}$\ Physik Department and Excellence Cluster Universe, Technische  Universit{\"a}t M{\"u}nchen, Germany\label{TUM} \\
$^{16}$\ Dipartimento di Fisica e Astronomia dell{`}Universit{\`a} di Padova, Padova, Italy\label{PDUNI} \\
$^{17}$\ INFN  Padova, Padova, Italy\label{PDINFN} \\
$^{18}$\ Physikalisches Institut, Eberhard Karls Universit{\"a}t T{\"u}bingen, T{\"u}bingen, Germany\label{TU} \\
$^{19}$\ Physik Institut der Universit{\"a}t Z{\"u}rich, Z{\"u}rich, Switzerland\label{UZH} \\
}

\begin{abstract}
Two neutrino double beta decay of \nuc{Ge}{76} to excited states of
\nuc{Se}{76} has been studied using data from Phase~I of the
\gerda\ experiment. An array composed of up to 14 germanium detectors
including detectors that have been isotopically enriched in \nuc{Ge}{76} was
deployed in liquid argon. The analysis of various possible transitions to
excited final states is
based on coincidence events between pairs of detectors where a de-excitation
\gadash{ray} is detected in one detector and the two electrons in the other.

No signal has been observed and an event counting profile likelihood analysis
has been used to determine Frequentist 90\,\% C.L. bounds for three
transitions: ${0^+_{\rm g.s.}-2^+_1}$: \thalftwo\unit{$>\baseT{1.6}{23}$}{~yr},
${0^+_{\rm g.s.}-0^+_1}$: \thalftwo\unit{$>\baseT{3.7}{23}$}{~yr} and
${0^+_{\rm g.s.}-2^+_2}$: \thalftwo\unit{$>\baseT{2.3}{23}$}{~yr}. These bounds
are more than two orders of magnitude larger than those reported
previously. Bayesian 90\,\% credibility bounds were extracted and
used to exclude several models for the ${0^+_{\rm g.s.}-0^+_1}$ transition.
\end{abstract}

%
\vspace{2pc}

\submitto{\JPG}
%
%
%
\section{Introduction}
\label{sec:intro}

The observation of neutrinoless double beta (\bb{0}) decay would imply physics
beyond the Standard Model of particle physics because the process manifests
lepton number violation.  Assuming light neutrino exchange as the dominant
process, the experimentally observable half-life would be connected to the
effective Majorana neutrino mass via a phase space factor ($F$) and a nuclear
matrix element (NME, ${\cal M}$). Calculations of ${\cal M}$ depend strongly
on the nuclear structure model chosen.  Uncertainties in the model translate
to uncertainties when converting a measured half-life into the Majorana
neutrino mass or its limit respectively~\cite{Suhonen:2012dn}.

Two neutrino double beta (\bb{2}) decay is a weak second order Standard Model
process which has been observed in eleven nuclides and double neutrino
electron capture ($2\nu$ECEC) in two nuclides with half-lives between
\thalftwo=\unit[$10^{18}-10^{24}$]{yr}~\cite{Tretyak:2002vc,Yang:2011wr}. Due
to the different reaction mechanisms the effective Majorana neutrino mass
\mbb\ enters only in the half-life for the \bb{0} mode, and not for the \bb{2}
mode:

 \begin{eqnarray}
 2\nu\beta\beta:\quad\quad  \left(T_{1/2}^{2\nu}\right)^{-1} 
      &=& F^{2\nu} \cdot \left|{\cal M}^{2\nu}\right|^2          \label{eq:2nbb}\\
 0\nu\beta\beta:\quad\quad   \left(T_{1/2}^{0\nu}\right)^{-1} 
      &=& F^{0\nu} \cdot \left|{\cal M}^{0\nu}\right|^2 \cdot \mbox{\mbb}^2
                                                              \label{eq:0nbb}
 \end{eqnarray}

The NME for \bb{2} and \bb{0} (${\cal M}^{2\nu}$, ${\cal M}^{0\nu}$) and the
respective phase space factors ($F^{2\nu}$, $F^{0\nu}$) are numerically
different but rely on similar model assumptions. Experimentally verifying the
calculations of Eq.~(\ref{eq:2nbb}) reduces to some extend uncertainties of
calculations of ${\cal M}^{0\nu}$. For a recent review see
Ref.~\cite{AvignoneIII:2008wm}.

Apart from decays into the ground state (g.s.), double beta decays can also
occur into excited states of the daughter nucleus. These decay modes are
expected to have a lower rate due to a smaller phase space. However, their
experimental signature is enhanced by the accompanying de-excitation
\gadash{rays}. Excited state transitions can in principle occur in both the
\bb{2} and the \bb{0} mode distinguishable only in the different sum of residual
electron energies.  These investigations provide additional information on the
nuclear structure models and more experimental constraints on the system of
Eqs.~(\ref{eq:2nbb} and \ref{eq:0nbb}).  So far only transitions to the first
excited $0^+_1$ state have been observed, first for
\nuc{Mo}{100}~\cite{Barabash:1995jt} in 1995 and for
\nuc{Nd}{150}~\cite{Barabash:2004vf} in 2004. Recent half-life values are
\thalftwo\unit{\baseT{=(7.5\pm1.2)}{20}}{\,yr}~\cite{Nemo3:2014iy} and
\thalftwo\unit{$=\baseT{(1.33^{+0.63}_{-0.36})}{20}$}{\,yr}~\cite{Barabash:2009fw},
respectively.

\begin{figure}[t]
\begin{center}
  \includegraphics[width=0.9\textwidth]{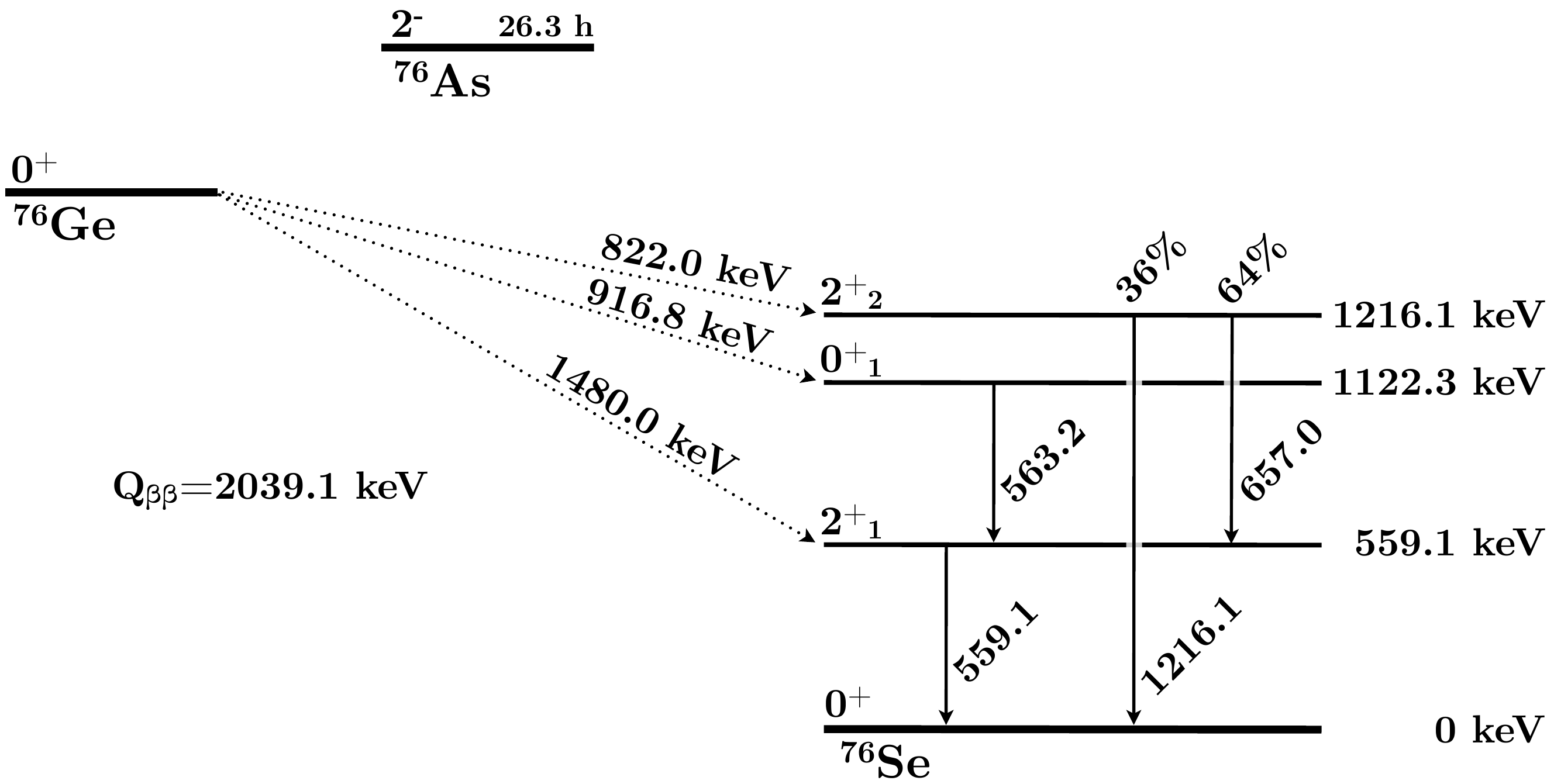}
  \caption{\label{fig:DecaySchemeGe76}
            Possible double beta decay modes of \nuc{Ge}{76} to excited states
            of \nuc{Se}{76} that are investigated in this work.
}
\end{center}
\end{figure}

The double beta decay of \nuc{Ge}{76} to the ground state of \nuc{Se}{76} has
a Q-value of \qbb=(2039.061$\pm$0.007)\,keV~\cite{Mount:2010fh} and can
potentially feed any $0^+$ or $2^+$ excited state in \nuc{Se}{76} up to this
energy. The search in this work focuses on the \bb{2} transitions from the
ground state $0^+_{\rm g.s.}$ of \nuc{Ge}{76} to the three lowest excited
states in \nuc{Se}{76}:
$                        
2^+_1$, $                
0^+_1$, and
$                        
2^+_2$ (see Fig.~\ref{fig:DecaySchemeGe76}).  
The phase spaces of \bb{2} transitions scale with the total available
energy~$E$ as $F^{2\nu}\sim E^{11}$ and reduce the rates for higher energetic
states. The rate is further suppressed by the spin-constraint for $2^+$
states. For the investigated transitions the existing experimental upper
limits for half-lives are shown in Table~\ref{tab:PreviousLimits}.  In
addition results from theoretical calculations based on various nuclear
structure models are presented. The largest rate is expected for the $0^+_{\rm
  g.s.}-0^+_1$ transition and lies within the experimental sensitivity of this
analysis.  Three calculations have been recently performed for this
transition. One is based on the renormalized proton-neutron QRPA using wave
functions from Ref.~\cite{Suhonen:2011dq}. This calculation predicts a
half-life of \thalftwo($0^+_1$)=\cpowten{(1.2-5.8)}{23}\,{yr} for axial vector
couplings of $g_A$=1.26 -- 1.00~\cite{PrivCommSuhonen}.  Employing a
microscopic interacting boson model (IBM-2), the half-life ratio between the
$0^+_{\rm g.s.}-0^+_1$ and the ground state transition $0^+_{\rm
  g.s.}-0^+_{\rm g.s.}$ is calculated with NMEs and phase space factors from
Refs.~\cite{Barea:2013eu,Kotila:2012bj}. The predicted half-life ratio is 3300
for \nuc{Ge}{76}~\cite{PrivCommIachello}. Scaling this ratio with our recently
measured ground state half-life of \thalftwo$(0^+_{\rm g.s.})$=$(1.926 \pm
0.095)\cdot 10^{21}$\,yr~\cite{Agostini:2015nwa} results in the predicted
excited state half-life \thalftwo($0^+_1$)=\cpowten{6.4}{24}\,yr independent
of $g_A$. A shell model (SM) calculation predicts
\thalftwo($0^+_1$)=\cpowten{(2.3-6.7)}{24}\,yr~\cite{PrivCommMenendez}
assuming $g_A$=1.26 -- 1.00. The range of \thalftwo($0^+_1$)\ additionally
encompasses results of two different effective interactions that have been
used also in Refs.~\cite{Menendez:2009dw} and~\cite{Caurier:2012ko}. The SM
calculation of \thalftwo($0^+_1$)\ to the ground state is in good agreement
with the experimental data. The IBM-2 and SM predictions are significantly
longer than other calculations. The current status for the experimental and
theoretical situation is summarized in Table~\ref{tab:PreviousLimits} for all
investigated transitions.

\begin{table}[t]
\begin{center}
\caption{\label{tab:PreviousLimits}
          Experimental half-life limits are compared to model predictions
          for \bb{2} excited state decay modes in \nuc{Ge}{76} as discussed in
          this paper. The energy of the final level is given.
          Abbreviations denote:  
                    HFB: Hartree-Fock-Bogoliubov,
                    QRPA: Quasi Random Phase Approximation,
                    MCM-QRPA: Multiple Commutator Model QRPA,
                    RQRPA: Renormalized proton-neutron QRPA,
                    IBM: Interactive Boson Model, 
                    SM: Shell Model.
}
\vspace*{2mm}
\footnotesize
\begin{tabular}{c|ccrc}
\toprule
\bb{2} decay mode & \thalftwo\ [yr]& model/exp. & ref.~ & year\\
\midrule
$0_{\rm g.s.}^+ - 2^+_1$ (559.1~keV) 
			  & \baseT{>6.3}{20} (\unit[68]{\,\%} C.L.)& exp. &~\cite{Beck:1992} & 1992\\ 
			  & \baseT{>1.1}{21} (\unit[90]{\,\%} C.L.)& exp.
&~\cite{Barabash:1995wr} & 1995\\ 
\cline{2-5}
			  & \baseT{1.2}{30} & SM &~\cite{Haxton:1984vu}  & 1984\\		 
			  & \baseT{5.8}{23} & HFB &~\cite{Dhiman:1994gq}  & 1994\\ 
			  & \baseT{5.0}{26} & QRPA &~\cite{Civitarese:1994fd}  & 1994\\ 
			  & \baseT{2.4}{24} & QRPA &~\cite{Stoica:1996np} & 1996\\ 
			  & \baseT{7.8}{25} & MCM-QRPA &~\cite{Aunola:1996hq} & 1996\\ 
			  & \baseT{1.0}{26} & RQRPA &~\cite{Toivanen:1997gz} & 1997\\ 			  
			  & \baseT{(2.4-4.3)}{26} & RQRPA &~\cite{Schwieger:1998cj} & 1998\\ 
			  & \baseT{2.0}{27} & RQRPA &~\cite{Unlu:2014ix} & 2014\\ 
\midrule
$0_{\rm g.s.}^+ - 0^+_1$ (1122.3~keV)
			  & \baseT{>6.3}{20} (\unit[68]{\,\%} C.L.)& exp. &~\cite{Beck:1992} & 1992\\ 
		 	  & \baseT{>1.7}{21} (\unit[90]{\,\%} C.L.)& exp. &~\cite{Barabash:1995wr} & 1995\\
			  & \baseT{>6.2}{21} (\unit[90]{\,\%} C.L.)& exp. &~\cite{Vasilev:2000zz} & 2000\\ 
\cline{2-5}
			  & \baseT{1.32}{21} & HFB &~\cite{Dhiman:1994gq}  & 1994\\ 
			  & \baseT{4.0}{22} & QRPA &~\cite{Civitarese:1994fd} & 1994\\
			  & \baseT{4.5}{22} & QRPA &~\cite{Stoica:1996np}  & 1996\\
			  & \baseT{7.5}{21} & MCM-QRPA&~\cite{Aunola:1996hq} & 1996\\
			  & \baseT{(1.0-3.1)}{23} & RQRPA &~\cite{Toivanen:1997gz} & 1997\\
			  & \baseT{(1.2-5.8)}{23} & RQRPA &~\cite{PrivCommSuhonen} & 2014\\
			  & \baseT{6.4}{24} & IBM-2 &~\cite{PrivCommIachello,Agostini:2015nwa} & 2014\\			  			 
			  & \baseT{(2.3-6.7)}{24} & SM &~\cite{PrivCommMenendez} & 2014\\			  			 
\midrule
$ 0_{\rm g.s.}^+ - 2^+_2$ (1216.1~keV)
			  & \baseT{>1.4}{21} (\unit[90]{\,\%} C.L.) & exp. &~\cite{Barabash:1995wr} & 1995\\
\cline{2-5}
			  & \baseT{1.0}{29} & QRPA &~\cite{Civitarese:1994fd} & 1994\\
			  & \baseT{1.3}{29} & MCM-QRPA &~\cite{Aunola:1996hq} & 1996\\
			  & \baseT{(0.7-2.2)}{28} & RQRPA &~\cite{Toivanen:1997gz} & 1997\\
\bottomrule
\end{tabular} 
\end{center}
\end{table}

\section{The coincidence analysis of \gerda\ Phase~I data}

The \gadash{cascade} following excited state transitions provides a well
defined experimental signature which enables large background suppression. The
granular installation of the \gerda\ setup is used to measure coincidences
between two germanium detectors. In the following the two detectors are
distinguished between (1) a ``source'' detector where the \bb{2} decay occurs
and the two electrons are detected and (2) a ``gamma'' detector where the
de-excitation \gadash{ray} is detected. This is achieved by searching for a
\gadash{ray} of interest in one detector and labeling the other one as
``source''. Note, that the distiction is made on the analysis level and that
it might not be unique in some rare cases.

\subsection{Event signature}

The signatures of the investigated transitions are listed in the following (see
also Fig.~\ref{fig:DecaySchemeGe76}).

$(i)$ The transition feeding the 559.1\,keV level ($0^+_{\rm g.s.}-2^+_1$) has
one single de-excitation \gadash{ray} of the same energy and thus a
$\beta\beta$ spectrum with 1480.0\,keV endpoint energy.

$(ii)$ The transition feeding the 1122.3\,keV level ($0^+_{\rm g.s.}-0^+_1$)
de-excites via the $2^+_1$ state. A \unit[563.2]{\,keV} \gadash{ray} is
followed practically immediately by the \unit[559.1]{\,keV} \gadash{ray}. The
angular correlation $W$ of the two \gadash{rays} in the $0_1^+ - 2^+_1 -
0^+_{\rm g.s.}$ cascade is given by $W(\theta) \propto 1-3\cos^2{\theta} +
4\cos^4{\theta}$ where $\theta$ is the angle between them~\cite{teubner}. The
analysis searches for one of the two \gadash{lines} in the ``gamma'' detector.
However, the energy resolution in \gerda\ does not allow to separate the two
\gadash{lines} and a single peak region is used as region of interest. The
$\beta\beta$ endpoint energy is reduced to 916.8\,keV.

$(iii)$ The transition to 1216.1\,keV level ($0^+_{\rm g.s.}-2^+_2$) has two
decay branches: branch~1 with two de-excitation \gadash{rays} via the $2^+_1$
state with energies of \unit[657.0]{\,keV} and \unit[559.1]{\,keV}. Branch~2
has a single de-excitation \gadash{ray} of \unit[1216.1]{\,keV} directly into
the ground state. The former has a branching ratio of \unit[64]{\,\%} leading
to \unit[36]{\,\%} for branch~2. The angular correlation $W$ of the two
\gadash{rays} in the $2_2^+ - 2^+_1 - 0^+_{\rm g.s.}$ cascade of branch~1 is
$W(\theta) \propto 1-\frac{15}{13}\cos^2{\theta} +
\frac{16}{13}\cos^4{\theta}$. The $\beta\beta$ endpoint is reduced to
\unit[822.0]{\,keV}. The two branches are treated separately in the analysis
and are later combined into a single value for the half-life.

\subsection{The \textsc{Gerda} experiment}

The GERmanium Detector Array (\gerda) is an experiment designed to investigate
neutrinoless double beta decay of \nuc{Ge}{76} with an array of high purity
germanium (HPGe) detectors made from material with the \nuc{Ge}{76} fraction
enriched to $\sim$87\,\%. The detectors are operated within a cryostat
containing 64\,m$^3$ of liquid argon (LAr) at the Laboratori Nazionali del
Gran Sasso (LNGS) of Istituto Nazionale di Fisica Nucleare (INFN). The
installation of the HPGe detectors in a closely spaced array with little
material between the detectors facilitates an anti-coincidence veto for the
\bb{0} search into the ground state suppressing \gadash{ray} background.

The \gerda\ setup is described in detail in Ref.~\cite{Ackermann:2013ey}.  The
array of HPGe detectors used in Phase~I of the experiment was organized into
4~strings of 3 to 5 detectors each.  Three different detector types were used:
semi-coaxial and BEGe detectors enriched in \nuc{Ge}{76} (\nuc{Ge}{\rm enr})
and semi-coaxial detectors with natural abundance (\nuc{Ge}{\rm nat}).  Five
\nuc{Ge}{\rm enr} semi-coaxial detectors are from the Heidelberg-Moscow
experiment~\cite{KlapdorKleingrothaus:2000sz}, three \nuc{Ge}{\rm enr}
semi-coaxial detectors from the \igex\ experiment~\cite{Aalseth:2002gf}, and
three \nuc{Ge}{\rm nat} semi-coaxial detectors from the GENIUS test
facility~\cite{KlapdorKleingrothaus:2002km}. All were refurbished for their
operation in LAr.  Additionally, an initial batch of five \nuc{Ge}{\rm enr}
BEGe detectors produced for \gerda\ were employed~\cite{bege7}. The detector
strings were lowered into LAr by a two-arm lock system supporting one string
in one arm and three strings in the other~\cite{Ackermann:2013ey}.

\subsection{Data sets and energy calibrations} 
The data of \gerda\ Phase~I are used in this analysis. As described in
Ref.~\cite{Agostini:2013kq}, some detectors were removed and replaced by the 5
BEGe detectors after an initial data taking period. A higher background was
observed for 49\,d due to this change (silver data set as defined in
Ref.~\cite{Agostini:2014dp}). For the present analysis this higher background
period is omitted. The periods before and after had different detector array
configurations and are treated independently in the analysis.

The data set is composed of runs of approximately 1 month each. Each selected
run shows a stable operation of the whole array~\cite{Agostini:2014dp}.  In
some runs, individual detectors are excluded from the analysis due to
temporary instabilities.  In the present analysis every detector - also if its
data is not used - is considered as source of the decay. A two-detector
coincidence may be registered also in case the decay occurs in an excluded
detector and the two \gadash{rays} are deposited in two active detectors. The
efficiency calculation is taking these possibilities properly into account.

In the analysis, the target mass is defined as constant within
each array configuration; changes due to inactive detectors are condensed into
the signal detection efficiency of the array. The live-time sums up to an
exposure of ${\cal E}$=31.04\,\kgyr\ for the whole data set including all
\nuc{Ge}{\rm nat} and \nuc{Ge}{\rm enr} detectors. The isotopic exposure of
\nuc{Ge}{76} is ${\cal E}_{76}$=22.3~\kgyr.

The same energy software-threshold of \unit[100]{\,keV} is applied to all
detectors. This basic threshold was applied to ensure a full reconstruction
efficiency for all detector energies.  After additional quality cuts and a
$\mu$-veto cut the data set contains 2710 two-detector events and 82
three-detector events. This can be compared with $\approx$\cpowten{7}{5}
single-detector events. The efficiency to detect de-excitation \gadash{rays}
in three-detector events is more than one order of magnitude smaller and thus
only two-detector events are further analyzed.

Energy calibrations were performed with \nuc{Th}{228} sources typically once
per week.  In case of an energy deposition in more than one detector,
cross-talk affects the reconstructed energy. Data taken by dedicated
calibrations and the \nuc{K}{42} \gadash{line} in the physics data allow to
measure the effect and hence to correct for it.  The energy dependence of the
cross-talk is linear in good approximation.  The exposure averaged energy
resolution (FWHM) of the \unit[583]{\,keV} \gadash{line} from \nuc{Tl}{208} is
\unit[4.2]{\,keV} for coincident events while it is \unit[3.8]{\,keV} for
events with energy depostition in one detector only. The uncertainty on the
resolution is estimated to be 10\,\%.

\subsection{ Monte Carlo simulations}

Monte Carlo (MC) simulations were used to construct a specific background
model for coincidence events and to determine the detection efficiencies.  The
background model described in Ref.~\cite{Agostini:2014dp} is used as a
starting point whereto additional background contributions had been added. Due
to the coincidence requirement the individual background contributions to the
spectra are of different significance when compared to the single-detector
spectrum. Therefore, the evaluation of the background sources was tuned
specifically for two-detector events.  The main sources of background are
contributions from \nuc{Bi}{214}, \nuc{Bi}{212}, \nuc{Tl}{208}, \nuc{Ac}{228},
\nuc{K}{40} and \nuc{Co}{60} on the detector holders, \nuc{K}{42} and
\nuc{Ar}{39} distributed homogeneously in the LAr, \nuc{K}{42} on the detector
n$^{+}$ contact and \nuc{Bi}{214} on the detector p$^+$ contact. Additionally,
one further background contribution from \nuc{Ag}{108m} in the signal cables
was considered which is not part of the minimal background model in
Ref.~\cite{Agostini:2014dp} and only visible in coincidence data.

\begin{figure}[h]
  \begin{center}
   \includegraphics[width=1.0\textwidth]{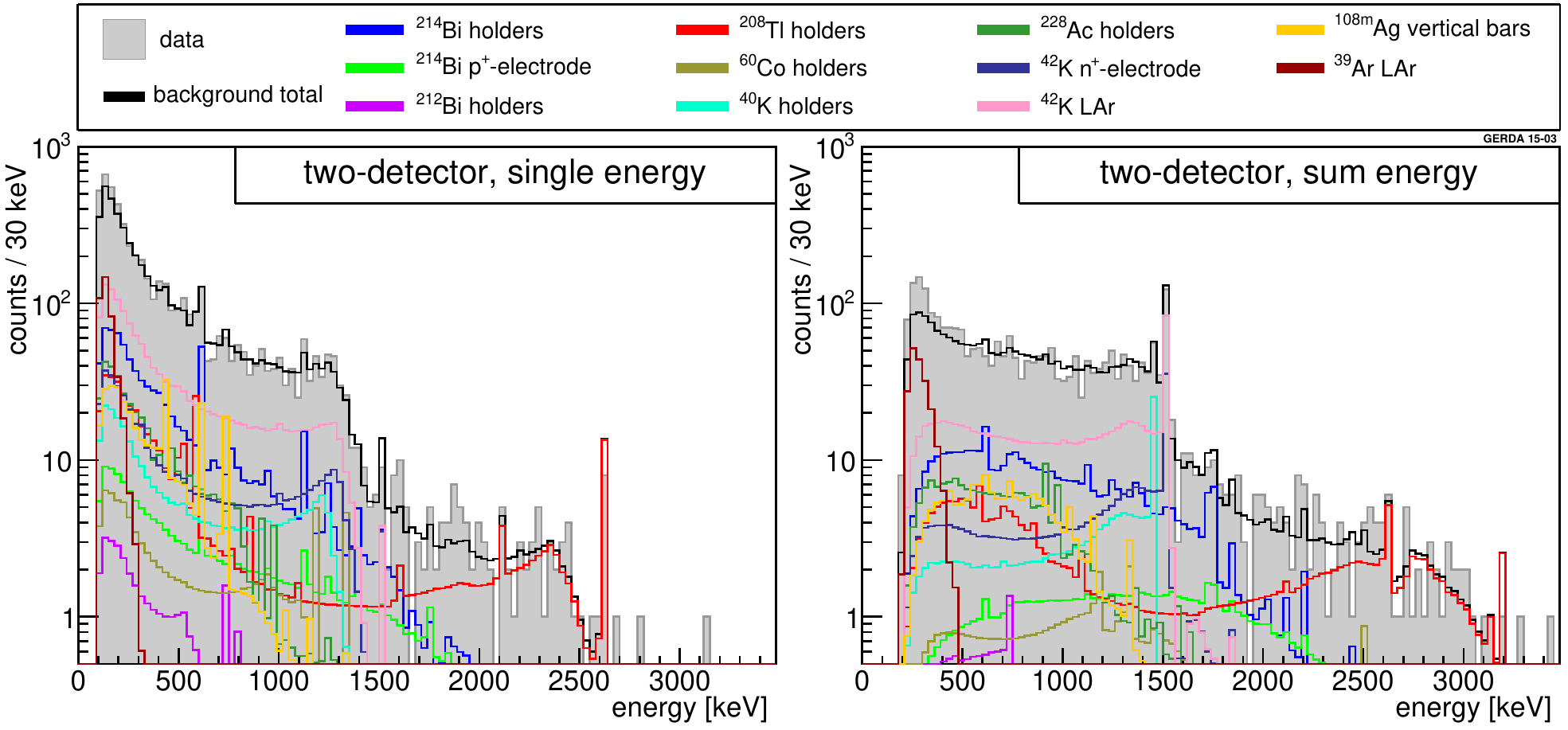}
   \caption{  \label{fig:pdf_ESpecBgComposition}
           Data and background model for two-detector events after an individual
           detector threshold of 100\,keV before cut optimization. In the
           single-energy  spectra (left) each detector has a separate entry
           per event in the histogram whereas in the sum-energy spectra
           (right) the two energies are summed before histogramming. Data
           events are shown in solid gray and the background components for
           the coaxial detectors by the colored lines.
}
\end{center}
\end{figure}

The energy spectra from the data obtained with all the detectors and from the
background model are shown in Fig.~\ref{fig:pdf_ESpecBgComposition} for
two-detector events.  This background description is used for the development
of cuts and the calculation of sensitivities but it does not enter itself in
the final analysis.  The left panel of Fig.~\ref{fig:pdf_ESpecBgComposition}
shows the individual detector energies in which each of the two detectors has
a separate entry. The right panel shows the sum-energy spectrum; i.e., the
total energy deposited in the array for a two-detector event. The shaded
histograms show the data which can be directly compared to the background
model shown in black. The individual background components for the
semi-coaxial detectors only are shown in color; due to low statistics the ones
for the BEGe detectors are omitted in the plot for clarity.

There are some differences found between background MC and data.  

At low energy beta decay events from \nuc{Ar}{39} with end point energy
565~keV dominate. The detected energy is highly sensitive to the exact
knowledge of the detector dead layer which is not available. Hence data and
Monte Carlo disagree here.  The probability of a two-detector coincidence due
to \nuc{Ar}{39} is small and highly sensitive to the exact thickness of the
detectors' dead layer. However, this effect is not relevant for the energy
regions investigated in this analysis.  Another excess is visible around
\unit[1.8]{MeV} in the single-energy spectrum in the data. However, it can not
be due to a missing \gadash{line} or beta spectrum in the background model and
furthermore the exess is not very significant.  The agreement between the
model and the data is sufficient for this analysis since the former is only
used for cut optimization~\cite{phd_thomas}.

MC simulations were used to determine the signal efficiency for each decay
mode. The simulations were performed with \mage~\cite{Boswell:hc} which
considers the angular correlation between the de-excitation \gadash{rays}.
The detectors that were inactive in a given run are also set as inactive in
the post-processing of the MC data. Each detector pair in each run has an
individual signal detection efficiency which is then life-time weighted into a
single number for the whole data set.
\section{Analysis}

The analysis is based on event counting in a region of interest (ROI) after a
sequence of cuts.  This is not a blind analysis, but special care was taken to
avoid biases.  The construction of the sequence of cuts starts with initial
choices that are then optimized to maximize sensitivity. This aims to prevent
\emph{ad hoc} choices of analysis parameters.  The MC background model and
efficiency are used for the cut optimization. Systematic uncertainties and
potential deficiencies in the background model only affect the choice of
analysis parameters and do not have a direct effect on the derived half-life
results since the background in the ROI is estimated from side bands (SB).
 
\begin{figure}[h]
        \centering
    	\includegraphics[width=1.0\textwidth]{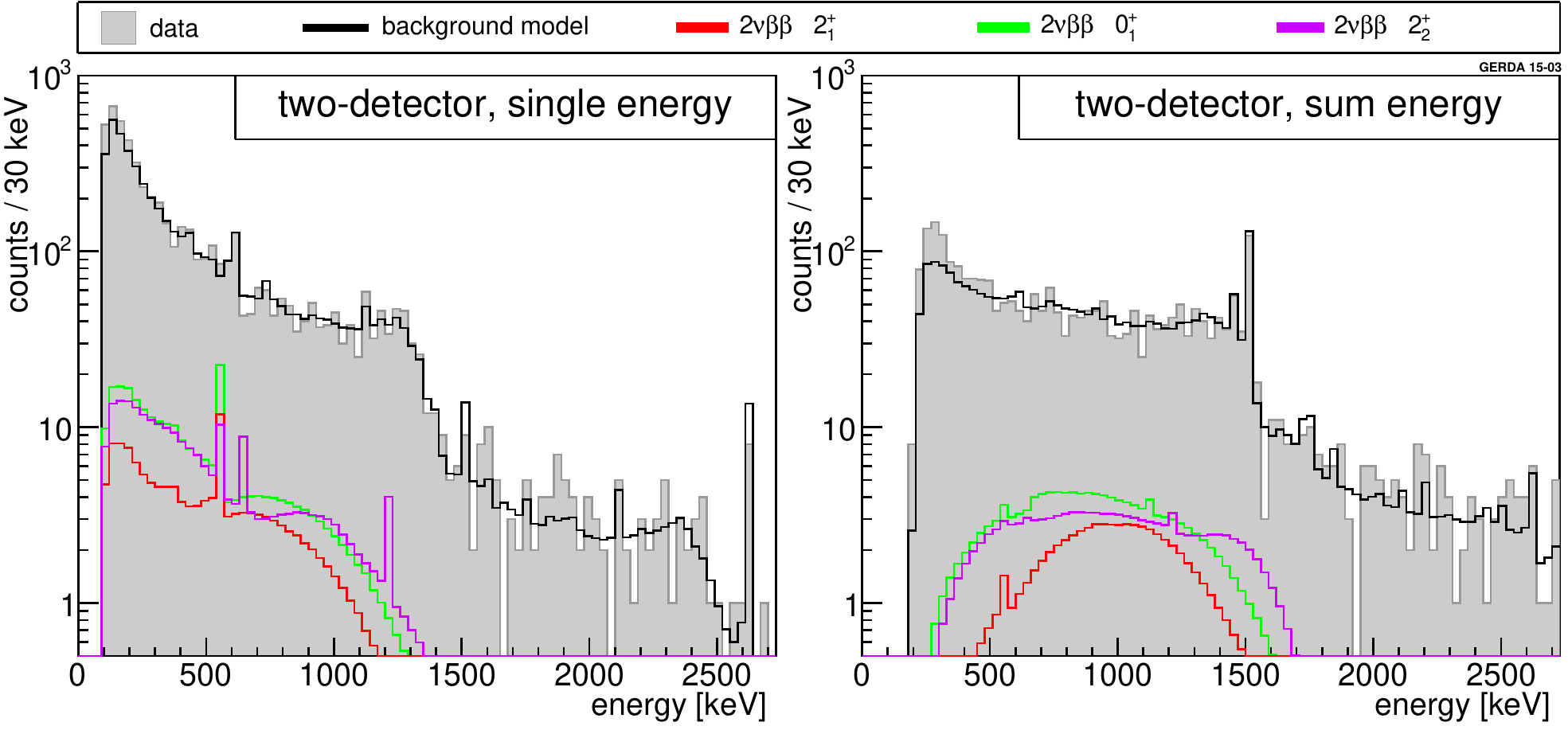}
        \caption{\label{fig:pdf_ESpecSignal}
            Illustration of the \bb{2} decays for two-detector events scaled
            to a half-life of \unit[\baseTsolo{23}]{\,yr} for each decay
            mode. Shown are the single-energy spectra (left) and the
            sum-energy spectra (right) befor cut optimization. Also shown are
            the background model (black line) and data (gray histogram).
}
\end{figure}

The spectra of the simulated \bb{2} decays scaled to \unit[\baseTsolo{23}]{\,yr}
half-life are shown in Fig.~\ref{fig:pdf_ESpecSignal}. Also shown is the
corresponding background model. The single-energy spectra and the sum-energy
spectra are shown separately. The \bb{2} decays have a continuous shape in the
sum-energy spectra due to the continuous electron component that is almost
always detected in the source detector. The single-energy spectra show
distinct de-excitation \gadash{lines} being detected outside the source
detector enabling strong background reduction.

\begin{figure}[h]
\begin{center}
  \includegraphics[width=0.47\textwidth]{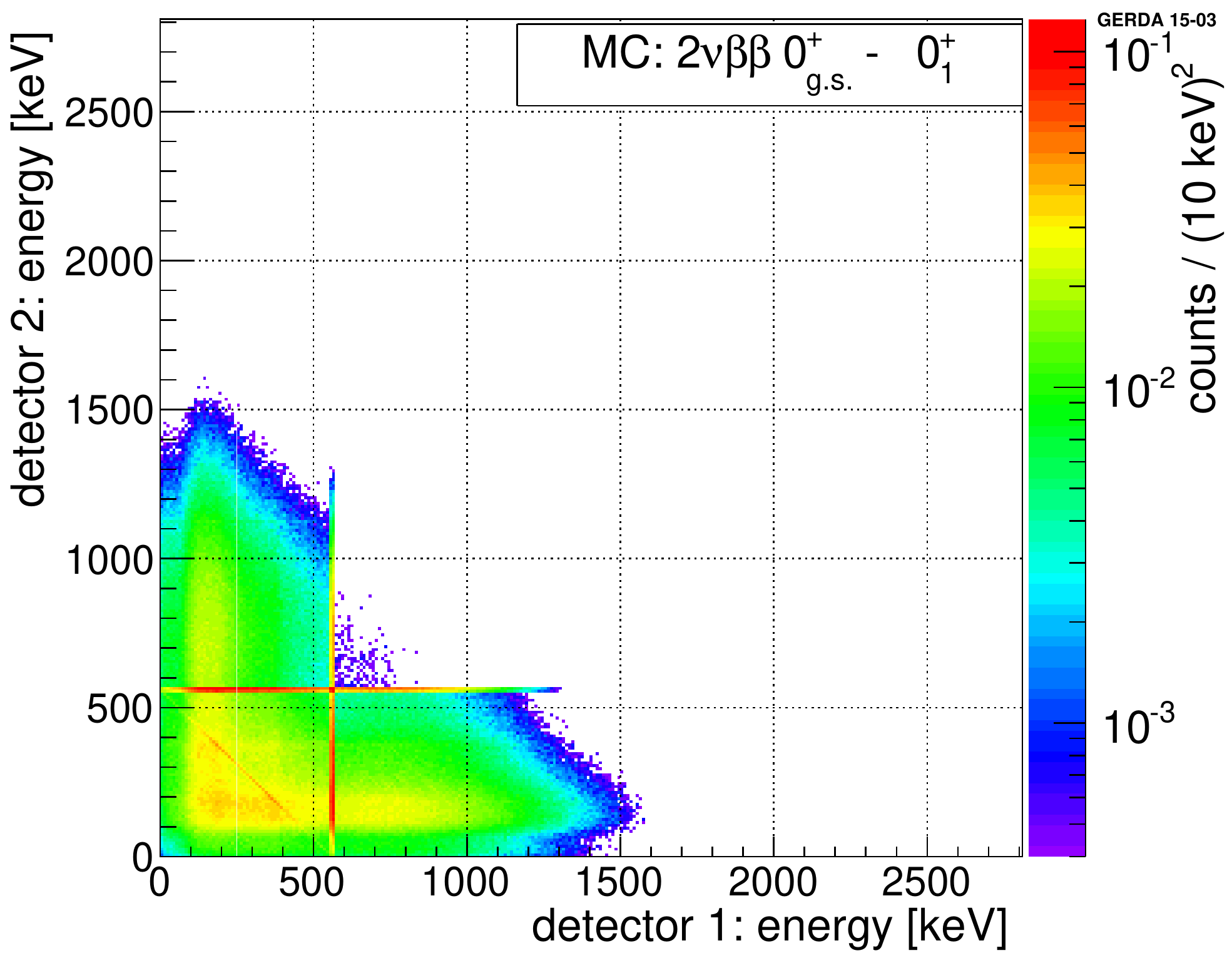}
\hspace*{1mm}
  \includegraphics[width=0.47\textwidth]{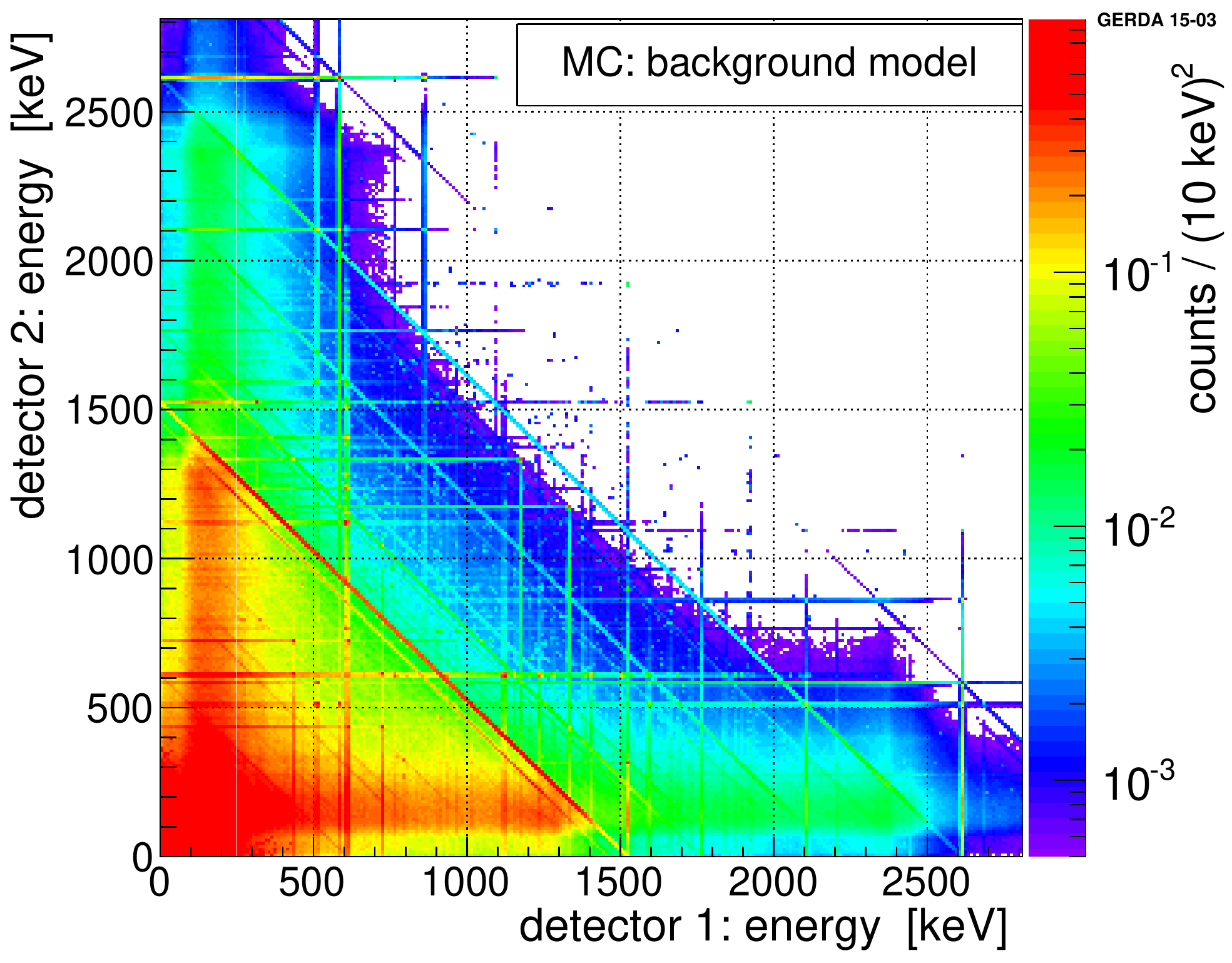}
\\
\parbox[b]{0.47\textwidth}{
   \includegraphics[width=0.47\textwidth]{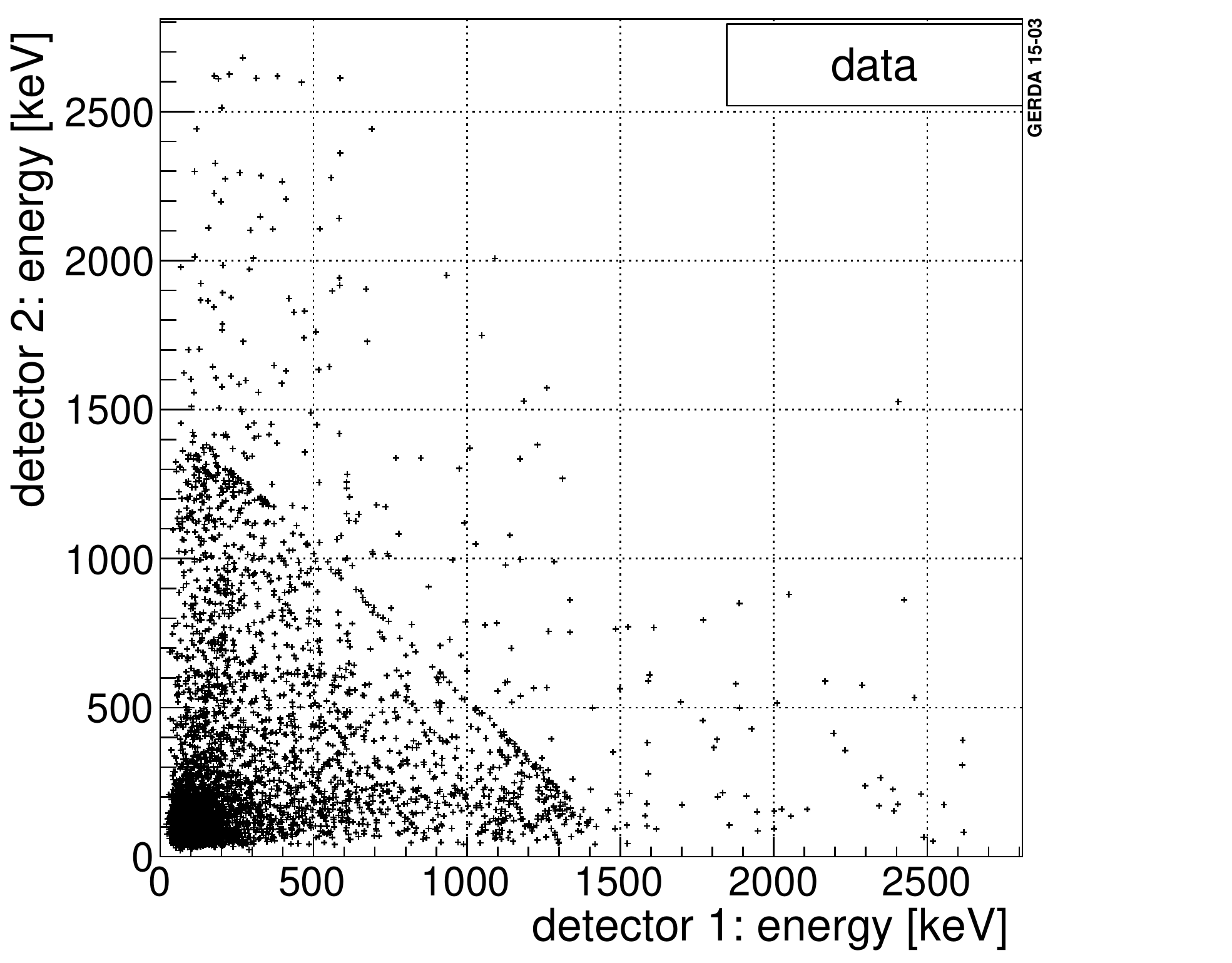}
}
\parbox[b]{0.47\textwidth}{%
\caption{\label{fig:pdf_scatter}%
         Scatter plot for coincident events showing the simulated \bb{2}
         signal process for the $0^+_{\rm g.s.} - 0^+_1$ decay mode scaled to
         \unit[\baseTsolo{23}]{\,yr} half-life (top left), the simulated
         background model (top right) and the data events (bottom left). The
         number of MC events is scaled to the Phase~I data set.
}
~\\[1mm]
}\parbox[b]{0.1\textwidth}{~}%
~\hfill
\end{center}
\end{figure}

The energy distributions of two-detector coincidence events are shown in
scatter plots (Fig.~\ref{fig:pdf_scatter}) for the simulated $0^+_{\rm
  g.s.}-0^+_1$ transition, the background model and the data.  The color scale
denotes the expected event densities for an excited state half-life of
\unit[\baseTsolo{23}]{\,yr} and for the expected number of background events
in the data set, respectively. The black points are data events.  In this
representation, horizontal and vertical lines are features of the
single-energy spectra whereas diagonal lines are features of the sum-energy
spectra.  Many high energy background \gadash{lines} originate from outside
the germanium detectors and can scatter into two detectors. These events can
be suppressed with a cut on the sum energy. In Fig.~\ref{fig:pdf_scatter} the
diagonal \gadash{line} from \nuc{K}{42} is clearly visible in the data and the
background model. Additional \gadash{lines} from \nuc{Tl}{208}, \nuc{K}{40}
and \nuc{Bi}{214} are only visible in the projections (see
Figs.~\ref{fig:pdf_ESpecBgComposition}).

\subsection{Sequence of cuts}

A sequence of four cuts is applied: 

\begin{enumerate}
\item[1.] {\bf standard cuts}: quality cuts and $\mu$ veto as in \gerda\ Phase~I
\item[2.] {\bf coincidence cuts}: specific two-detector cuts optimized for each
  decay mode 
\item[3.] {\bf background cuts}: exclusion of background \gadash{lines} 
\item[4.] {\bf detector pair cuts}: cuts to select detector pairs that exhibit
  a high efficiency
\end{enumerate}

The cuts are designed to optimize the sensitivity~$\widehat{T_{1/2}}$,
expressed by the figure of merit $S = \epsilon/\sqrt{B}$ with the signal
detection efficiency $\epsilon$ and the number of surviving background
events~$B$.  The efficiency and the background in the sensitivity study are
entirely based on MC simulations~\cite{Lehnert:2015}.

The {\bf standard cuts} include quality cuts and the muon veto cut and are
initially applied to the data set. They are identical to the cuts in the
\gerda\ \bb{0} analysis (without pulse shape cuts) which are described in
Ref.~\cite{Agostini:2013kq} and references therein.

The {\bf coincidence cuts} are specific for each decay mode. They require a
coincident event with either of the two detectors having the full energy of
any de-excitation \gadash{ray} within a peak (energy) window size (PWS) of
\unit[$\pm 2$]{$\sigma_{\rm E}$} where $\sigma_{\rm E}$ denotes the energy
resolution at energy $E$.  The PWS defines the specific ROI and SBs.
Furthermore, a sum energy limit of \unit[2039]{\,keV} is applied to exclude
events with a total energy deposition larger than the possible energy release
in \nuc{Ge}{76}.

SBs are created by shifting the energy window of the coincidence cuts to lower
and higher energies. The window size of a single SB is the same as for the
corresponding ROI. To reduce the statistical uncertainty in the background
estimation, a total of 4 SBs are defined for each decay mode; two at lower
energies and two at higher energies.  Based on the MC background model the SB
regions are chosen for each decay mode individually. They are as close as
possible to the ROI and avoid background \gadash{lines} in the single-energy
spectra.

For two-detector events it is possible that the same event is tagged in the
ROI and in a SB. A signal event that is tagged in the ROI has a 4\,\% chance
to be tagged also in one of the SBs. A background event in a SB has a 2\,\%
chance to be tagged also in another SB or the ROI. Events that are tagged in
more than one of these five regions are rejected to avoid double counting in
the statistical analysis. This reduces the signal efficiency by 4\,\% and the
background by 2\,\%~\cite{Lehnert:2015}.  One event in the ROI of $0^+_{\rm
  g.s.}-2^+_2$ is removed by this requirement.

\begin{figure}[h]
        \centering
 \includegraphics[width=0.7\textwidth]{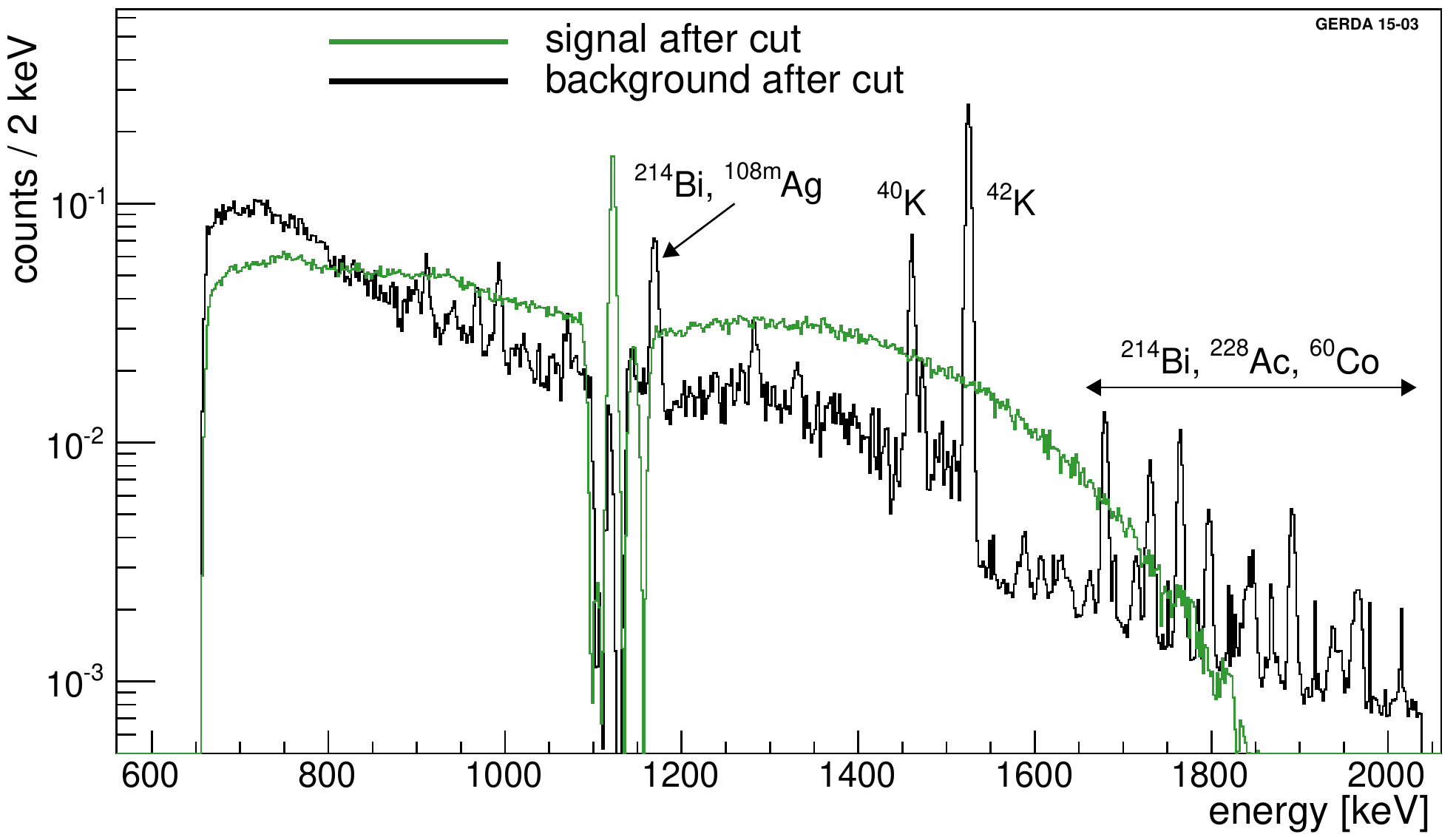}
        \caption{\label{fig:eSpecAfterCut1}
              Simulated sum-energy spectrum after coincidence cuts for the
              $0^+_{\rm g.s.}-0_{1}^{+}$ transition. The background model
              (black) and signal process scaled to \unit[\baseTsolo{23}]{\,yr}
              half-life (green) are shown.
}
\end{figure}

The {\bf background cuts} are motivated by background \gadash{lines} that
interfere with the ROIs. Fig.~\ref{fig:eSpecAfterCut1} shows the simulated sum-energy spectrum of the background model and the signal process after the
coincidence cut for the $0^+_{\rm g.s.}-0^+_1$ transition.  The low side
cutoff is created by the smallest \gadash{ray} energy of interest
(\unit[559.1]{\,keV}) in one detector and the single detector threshold of
\unit[100]{\,keV} in the other.  The high cutoff is given by the sum energy
limit.  The dips in the spectra are created by rejecting events that are
tagged in the ROI and a SB or two SBs simultaneously.  The peak at
\unit[1122.3]{\,keV} for the signal process (green) is created by decays that
occur in the dead layer of a detector or inside an excluded detector. In this
case the $\beta$ energy of the event is not detected and the two \gadash{rays}
trigger a two-detector event with discrete sum energy.

The strongest peaks in the background spectrum (black line in
Fig.~\ref{fig:eSpecAfterCut1}) are found at \unit[1524.7]{\,keV} and
\unit[1460.8]{\,keV} belonging to \nuc{K}{42} and \nuc{K}{40},
respectively. They represent the region in the scatter plot
(Fig.~\ref{fig:pdf_scatter}, bottom) where the diagonal sum energy
\gadash{lines} cross the horizontal and vertical single energy lines of the
\bb{2} decay. The next strongest feature in the background model is the peak
around \unit[1170]{\,keV} which is a combined structure: Firstly, a
\unit[609.3]{\,keV} \nuc{Bi}{214} \gadash{ray} coincides with another
\nuc{Bi}{214} \gadash{ray} that scatters and deposits energy around the ROI
of \unit[559.1]{\,keV} in one detector; the \unit[609.3]{\,keV} \gadash{ray}
is fully detected in the other detector creating a sum energy around
\unit[1170]{\,keV}. Secondly, a similar coincidence with a \unit[614.3]{\,keV}
\nuc{Ag}{108m} \gadash{ray} and another one from the \nuc{Ag}{108m} decay can
happen. The other peaks originate mainly from \nuc{Bi}{214}, \nuc{Ac}{228} or
\nuc{Co}{60}. They are created either (1) by a single \gadash{ray} scattering
into two detectors or (2) by two \gadash{rays} with one \gadash{ray} fully
detected in one detector. In both cases an energy around \unit[$\approx
  560$]{\,keV} has to be deposited in at least one detector. This can be
identified in Fig.~\ref{fig:pdf_scatter} where the horizontal/vertical cut
window crosses anti-diagonal background lines in scenario (1) or
horizontal/vertical background lines in scenario (2).

The background cuts are applied to all decay modes in the same
way. \nuc{K}{42} and \nuc{K}{40} are excluded by requiring the detector sum
energy not to be in the range of \unit[$1524.7\pm 5.5 $]{\,keV} nor in
\unit[$1460.8\pm 5.4$]{\,keV}. The energy ranges correspond to
2$\sigma_{\rm E}$. Additionally, the individual detector energy is required to
be outside \unit[$611.0\pm5.7$]{\,keV} to exclude the \nuc{Bi}{214} and
\nuc{Ag}{108m} background.

At this stage, the combination of coincidence cuts and background cuts can be
optimized in three ways. The individual detector threshold is scanned from
\unit[100]{\,keV} to \unit[500]{\,keV} in steps of \unit[50]{\,keV}.  This is
equivalent of increasing the low energy cutoff in
Fig.~\ref{fig:eSpecAfterCut1}.  The figure of merit $S$ is determined for each
step and the individual detector threshold with the highest $S$ is chosen as
the optimized cut. In a similar way the sum energy limit is scanned from
\unit[1000]{\,keV} to \qbb\ in steps of \unit[50]{\,keV}. This is equivalent
of changing the high energy cutoff in Fig.~\ref{fig:eSpecAfterCut1}. Finally
also the peak window size is scanned from \unit[1]{$\sigma$} to
\unit[2]{$\sigma$} energy resolution in steps of \unit[0.2]{\,keV}. The
optimization of these cut parameters is performed individually for all three
decay modes. The optimal values are shown in
Table~\ref{tab:SummaryOptimization} along with their relative sensitivity gain
compared to the base values.

\begin{table}[h]
\begin{center}
\caption{\label{tab:SummaryOptimization}
          Optimized values for individual detector threshold (IDT), sum energy
          limit (SEL) and the peak window size (PWS). The respective ROIs are
          given as absolute ranges; the $^\star$ indicates an expanded range
          due the occurrence of more than one $\gamma$ line. The sensitivity
          gain is related to the base values before background cuts.
}
\begin{tabular}{l|rrrr|r}
\toprule
mode & IDT  & SEL & PWS & size ROI & sensitivity gain \\
	 & [keV] & [keV] & [keV] & [keV] & [ \%]\\
\midrule
$0^+_{\rm g.s.}-2_{1}^{+}$ & 450 & 1750 & $\pm2.6$ & $[556.5-561.7]$ &  41.3 \\
$0^+_{\rm g.s.}-0_{1}^{+}$ & 250 & 1750 & $\pm2.2$ & $[556.9-565.4]^\star$ &  18.9 \\
$0^+_{\rm g.s.}-2_{2}^{+}$ B1 & 250 & 1800 & $\pm2.4$ & $[556.7-561.5]$  &  20.4 \\
 &  &  &$\pm3.0$ & $\lor [654.0-660.0]^\star$ &  \\
$0^+_{\rm g.s.}-2_{2}^{+}$ B2 & 300 & 1850 & $\pm2.8$ & $[1213.3-1218.9]$ &  98.5 \\
\midrule
base values & 100 & 2039 & $2\sigma_E$ & &\\
\bottomrule
\end{tabular}
\end{center}
\end{table}

The {\bf detector pairs} of coincidence events which are used in the analysis
are also tuned to maximize $S$. Each detector pair is composed of the gamma
detector and the source detector which are identified by containing the energy
deposition in the ROI or an arbitrary energy deposition within the allowed
energy range, respectively. This results in an asymmetric effect for
\nuc{Ge}{\rm nat} and \nuc{Ge}{\rm enr} detectors: \nuc{Ge}{\rm enr} detectors
are more likely to be the source of a \bb{2} decay than \nuc{Ge}{\rm nat}
detectors but have an equal chance to detect \gadash{rays}. Taking into
account all data sets there are in
total 28 detector pairs in the \gerda\ array and 56 detector pairs if the
distinction between source detector and gamma detector is taken into account.
The maximization of $S$ selects 37 pairs. The sensitivity gain
from the pair selection is between \unit[7 and 10]{\,\%} depending on the decay
mode. More detailed information on the sequence of cuts are reported in
Ref.~\cite{Lehnert:2015}.

After applying the full sequence of cuts, the validity of the SB regions is
tested.  For these regions a flat background is required.  In case the SBs are
symmetrically placed around the ROI, also a linear background is sufficient.
Table~\ref{tab:SidebandSummary} shows the expected and the observed event
count for each SB. For the $2_{1}^{+}$ mode the model predicts a flat
background while for the others the expected counts at higher energies (SB4)
is slightly smaller. However the average agrees with the background model
prediction in the ROI.  This validates the SBs to be used as background
estimators for the ROI.

\begin{table}[h]
\begin{center}
\caption{\label{tab:SidebandSummary}
          Summary of SBs after the complete sequence of cuts for each decay
          mode. Listed are the relative position $\Delta E$ of the four SBs
          compared to the ROI, the number of events N$_{\rm MC}$ in the MC
          background model and the number of events N$_{\rm data}$ in the data
          set.  Additionally the background expectations in the ROI and the
          observed events are shown after application of all
          cuts. The uncertainty of the MC expectations denote statistical
          uncertainties only.
}
\begin{tabular}{lrrr||lrrr}
\toprule
Region & $\Delta E$ & N$_{\rm MC}$ & N$_{\rm data}$ & Region & $\Delta E$ & N$_{\rm MC}$ & N$_{\rm data}$\\
\toprule
\multicolumn{ 4}{l||}{$0_{\rm g.s.}^+ - 2_{1}^{+}$  \unit[$559.1$]{keV}} & \multicolumn{ 4}{l}{$0_{\rm g.s.}^+ - 0_{1}^{+}$ \unit[$559.1\ \&\ 563.2$]{keV}} \\
\midrule
 SB 1 & \unit[$-7.5$]{keV} &  $2.5\pm 0.1$ & 2  &   SB 1 & \unit[$-12$]{keV} &  $8.0\pm 0.1$ & 7\\
 SB 2 & \unit[$+7.5$]{keV} &  $2.4\pm 0.1$ & 0  &   SB 2 & \unit[$+12$]{keV} &  $7.5\pm 0.1$ & 11\\
 SB 3 & \unit[$-15$]{keV}  &  $2.6\pm 0.1$ & 3  &   SB 3 & \unit[$-24$]{keV} &  $8.3\pm 0.1$ & 7\\
 SB 4 & \unit[$+15$]{keV}  &  $2.4\pm 0.1$ & 5  &   SB 4 & \unit[$+35$]{keV} &  $6.8\pm 0.1$ & 9\\
 average SB &              &  $2.5\pm 0.1$ & 2.5&   average SB &             &  $7.7\pm 0.1$ & 8.5\\
 ROI  &                    &  $2.5\pm 0.1$ & 2  &   ROI  &                   &  $7.9\pm 0.1$ & 5\\
\midrule
\midrule
\multicolumn{ 4}{l||}{$0_{\rm g.s.}^+ - 2_{2}^{+}$ branch~1: \unit[$559.1\ \&\ 657.0$]{keV}} & \multicolumn{4}{l}{$0_{\rm g.s.}^+ - 2_{2}^{+}$ branch~2: \unit[$1216.1$]{keV}}\\
\midrule
 SB 1 & \unit[$-8$]{keV}  &  $8.5\pm 0.1$ & 6   &  SB 1 & \unit[$-19$]{keV} &  $0.52\pm0.02$ & 1\\
 SB 2 & \unit[$+18$]{keV} &  $8.1\pm 0.1$ & 5   &  SB 2 & \unit[$+10$]{keV} &  $0.39\pm0.02$ & 0\\
 SB 3 & \unit[$-16$]{keV} &  $8.7\pm 0.1$ & 6   &  SB 3 & \unit[$-27$]{keV} &  $0.58\pm0.02$ & 0\\
 SB 4 & \unit[$+35$]{keV} &  $7.9\pm 0.1$ & 12  &  SB 4 & \unit[$+47$]{keV} &  $0.32\pm0.02$ & 1\\
 average SB &             &  $8.3\pm 0.1$ & 7.25&   average SB &            &  $0.45\pm0.01$ & 0.5\\
 ROI  &                   &  $8.3\pm 0.1$ & 6   &   ROI  &                  &  $0.40\pm0.02$ & 0\\
\bottomrule
\end{tabular}
\end{center}
\end{table}

\section{Results}

\begin{figure}[Ht]
\begin{center}
 \includegraphics[width=0.48\textwidth]{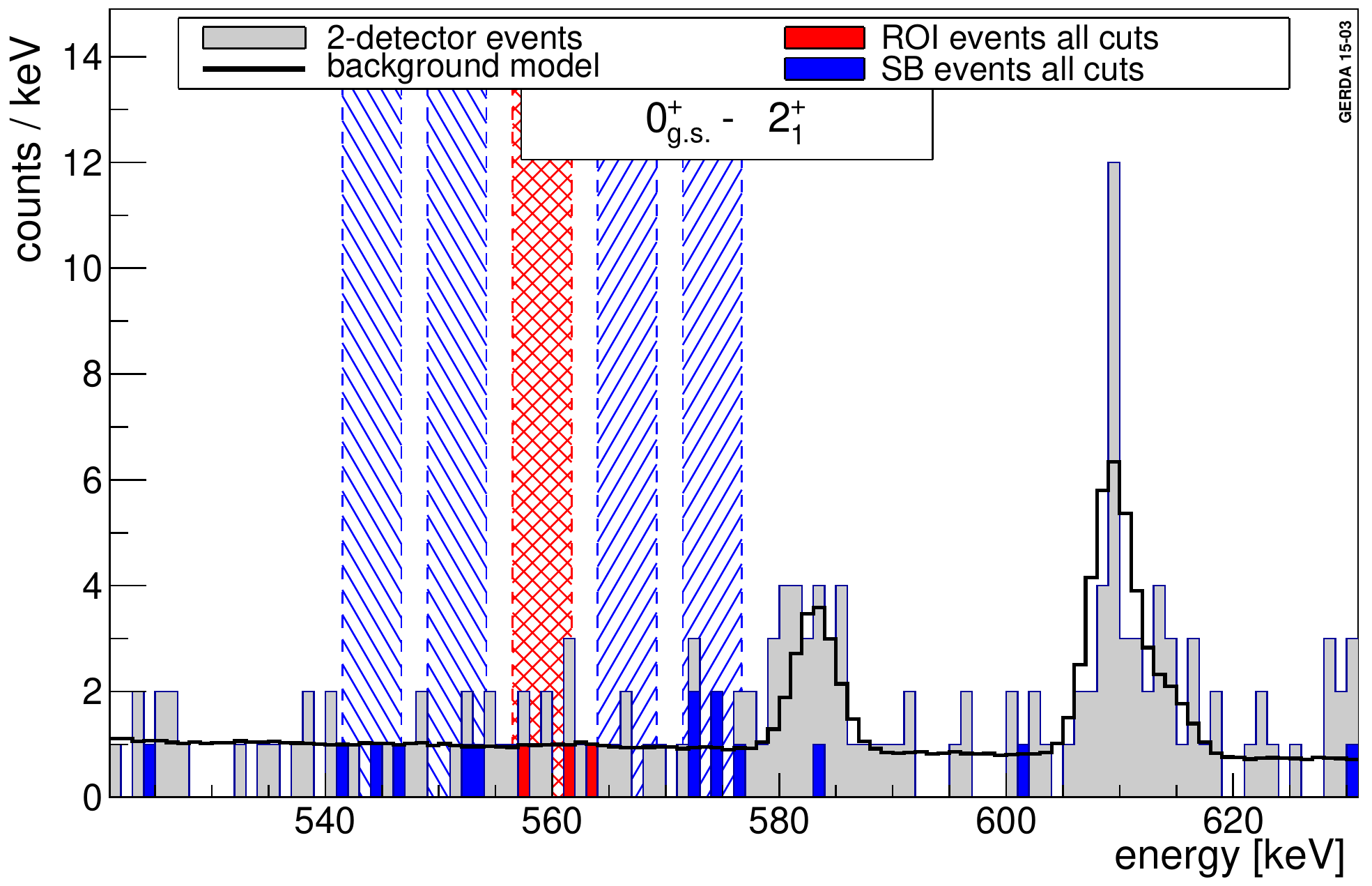}
\hspace*{1mm}
 \includegraphics[width=0.48\textwidth]{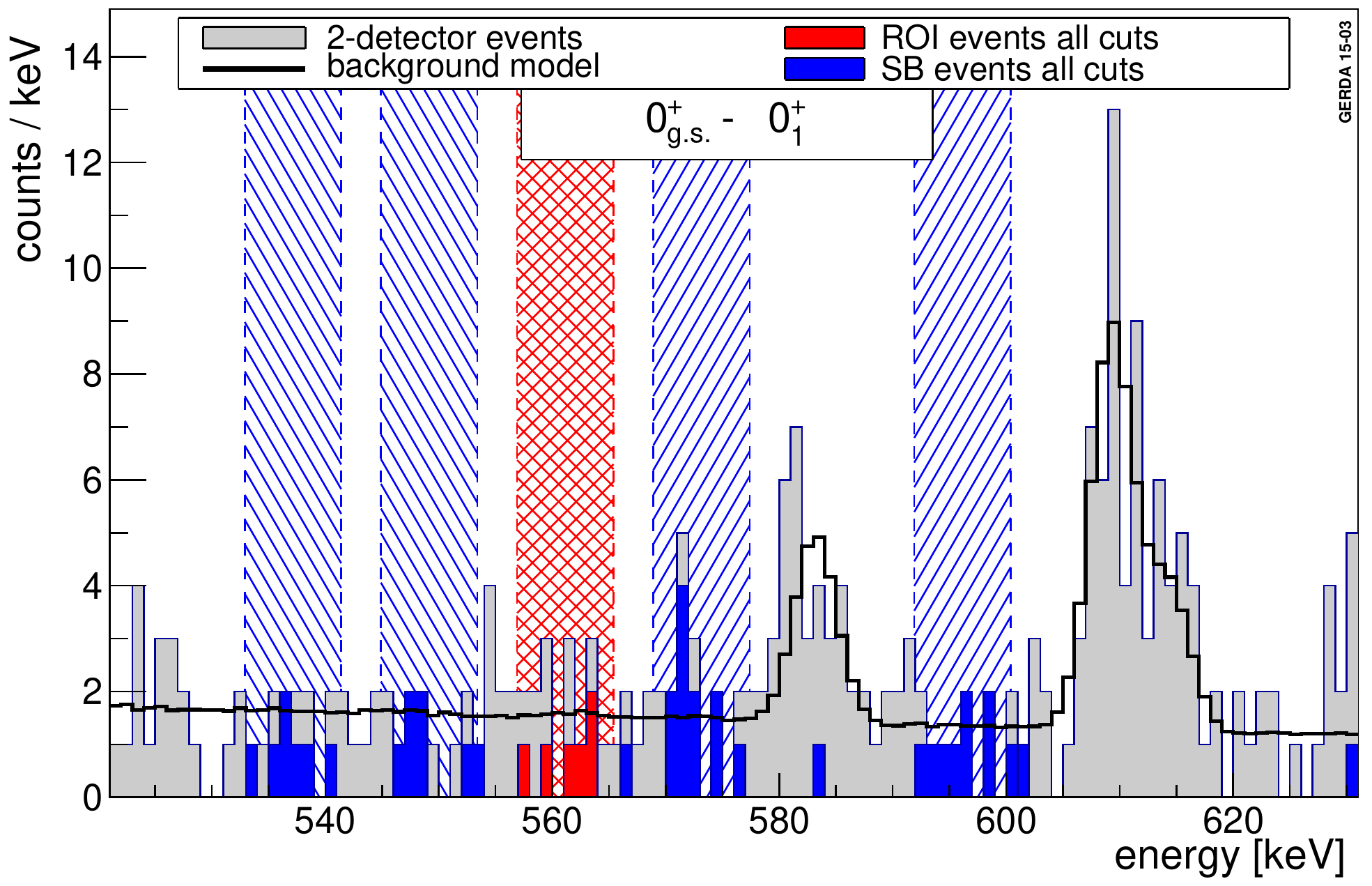}
~\\
 \includegraphics[width=0.48\textwidth]{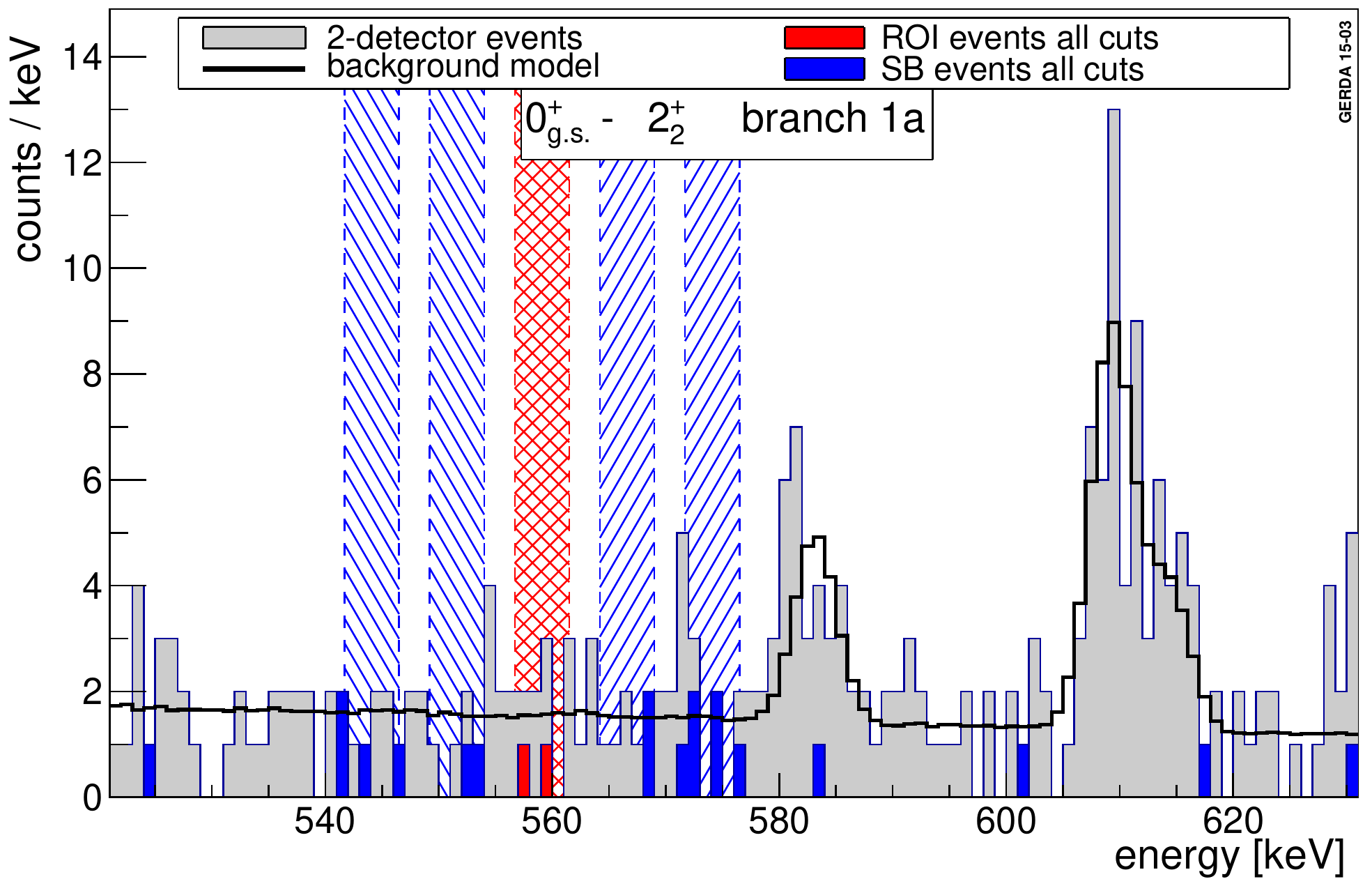}
\hspace*{1mm}
 \includegraphics[width=0.48\textwidth]{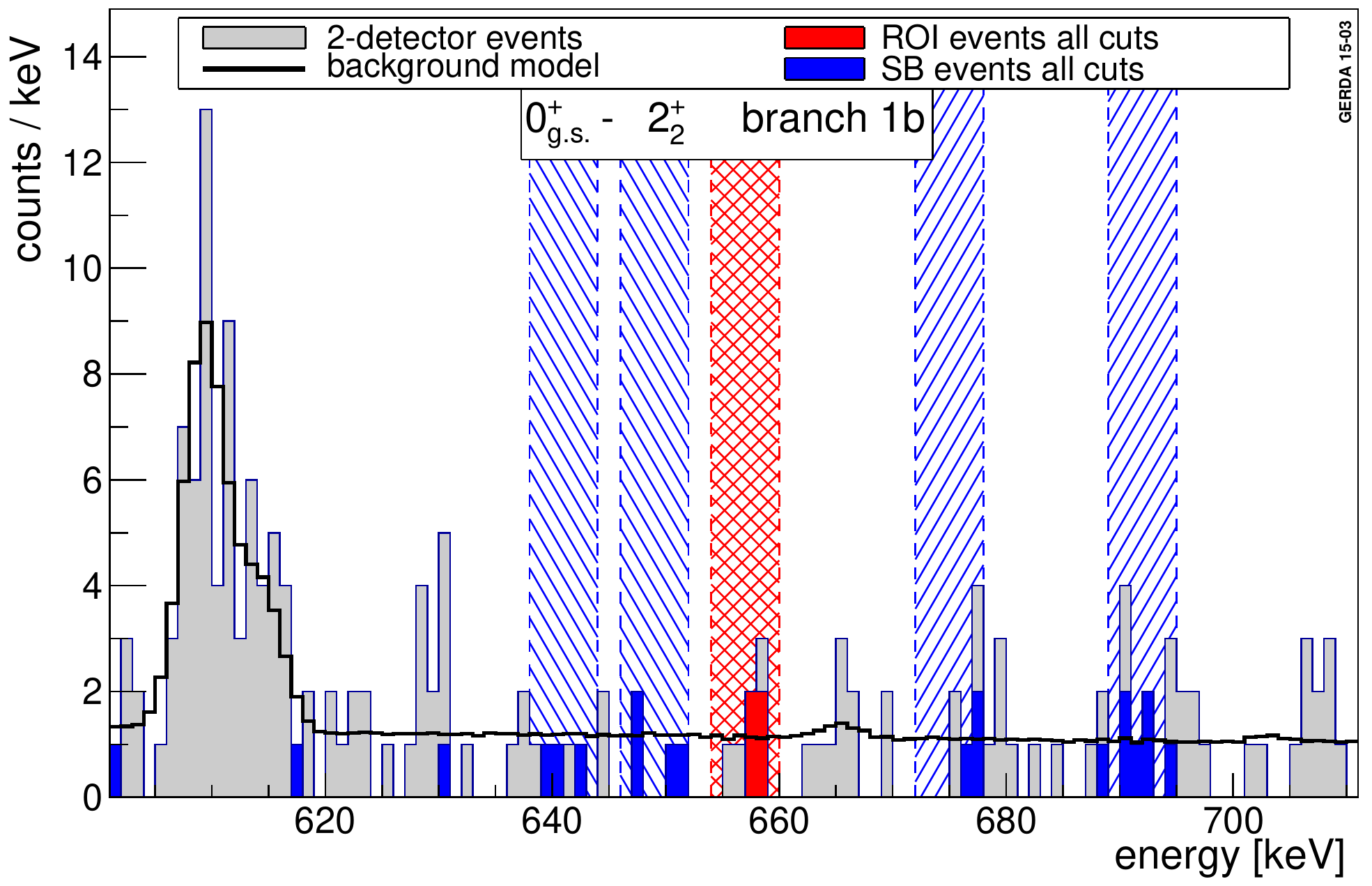}
~\\
 \includegraphics[width=0.48\textwidth]{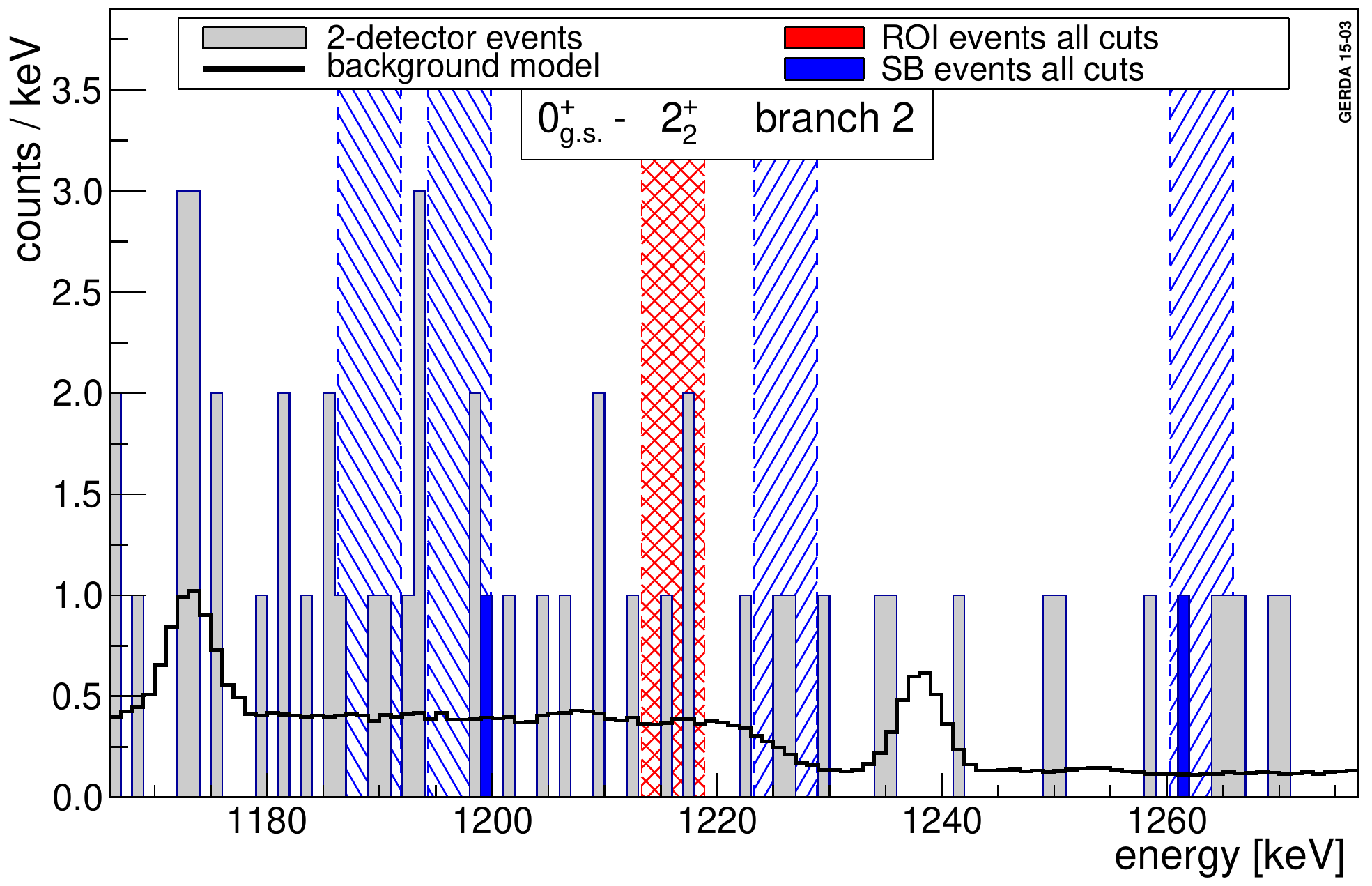}
\hspace*{1mm}
 \caption{\label{fig:pdf_SB_2nubb}
              Single-energy spectra around the respective ROI for all decay
              modes. Shown 
              are all two-detector events for the optimized individual detector
              threshold and sum energy limit (light gray) and the
              corresponding background curves (black). The optimized cuts
              result in different two-detector spectra for each decay
              mode. Also shown are the ROI (shaded red) and SB region (shaded
              blue). Highlighted are events that are tagged as ROI (red) and
              SB (blue) after all cuts and that are used for the limit
              setting. Note that the histograms contain two entries per event
              and that one entry may lie outside the tagging region.
}
\end{center}
\end{figure}

The single-energy spectra around the respective ROI are shown in
Fig.~\ref{fig:pdf_SB_2nubb} for all decay modes. All two-detector coincident
events with decay mode optimized individual detector threshold and sum energy
limit are shown in light gray; no other cuts are applied. The corresponding MC
background model is shown in black for illustration.  Events passing all cuts
have exactly one entry in one of the 5 intervals (but two entries in the
single-energy spectrum).  The ones in the ROI are shown in red and the ones of
the SBs in blue.

Frequentist \unit[90]{\%} confidence level and Bayesian \unit[90]{\%}
credibility lower values were calculated for $T_{1/2}$. The Frequentist values
were based on a bounded profile likelihood test statistic~\cite{Rolke:2005elb}
increased by 2.7 compared to the minimum. It was verified that this method
has always sufficiency coverage. For the Bayesian credibility limit, a flat
prior in $T^{-1}_{1/2}$ was assumed as well as a flat prior in the background
level.

In both approaches the same likelihood was used which is constructed for the
inverse half-life $T^{-1}_{1/2}$. In case of two decay branches as for \bb{2}
$0^+_{\rm g.s.} - 2^+_2$, the likelihood is treated as two individual data
sets with a common $T^{-1}_{1/2}$. The expectation for the signal counts for a
given decay branch~($k$) is:
\begin{eqnarray}
s_k = \ln{2} \cdot \eta_k \cdot {\cal E} \cdot T_{1/2}^{-1},
\end{eqnarray}
with the decay branch specific efficiency $\eta_k$. The efficiency is the
live-time weighted averaged detection efficiency of a \bb{2} decay excited
state event over all runs in the two array configurations of \gerda\ 
Phase~I. The exposure ${\cal E}$ is defined as the
combined total isotopic exposure for \nuc{Ge}{76} in the data set for all
detectors. The expected number of events in the ROI is:
\begin{eqnarray}
\mu_k = \frac{b_k}{4} + s_k
\end{eqnarray}
using the total background from all 4 SBs $b_k$. 

The full expression of the likelihood is constructed with three terms: (1) a
Poisson term describing the probability of signal plus background in the ROI,
(2) a Poisson penalty term accounting for the uncertainty of the background
level in the ROI and (3) a Gaussian penalty term accounting for all systematic
uncertainties condensed into the efficiency. The likelihood depends on
$T^{-1}_{1/2}$, the number of counts in the ROI ($n_k$), the number of counts
in all SBs ($m_k$) and the efficiency expectation ($\epsilon_k$):

\begin{eqnarray}
\label{eq:likelihood}
\mathcal{L}\left(n_k,m_k,\epsilon_k|T_{1/2}^{-1},b_k,\eta_k\right)\\ \nonumber
  = \prod_{k}
   \left[ \frac{(\mu_k)^{(n_k)}}{(n_k)!}  \cdot e^{-\mu_k} \right] 
  \cdot \left[ \frac{(b_k)^{(m_k)}}{(m_k)!} \cdot  e^{-b_k} \right] 
  \cdot \left[ \frac{1}{\sigma_{\epsilon_k} \sqrt{2\pi}}
   \cdot  e^{-\frac{1}{2}\left(\frac{\epsilon_k-\eta_k}{\sigma_{\epsilon_k}}\right)^2} \right].
\end{eqnarray}

For the profile likelihood we calculate $-2 \log{\mathcal{L}}$.  In the
extraction of the posterior probability~\cite{Caldwell:2009kh}, flat priors
for all fit parameters $T_{1/2}^{-1}$, $b_k$ and $\eta_k$ were used since the
prior information on the background and efficiency is included in the
likelihood with penalty terms.

Systematic uncertainties on the signal efficiency are estimated with MC
simulations and combined into a single value $\sigma_{\epsilon_k}$ assuming no
correlations.  The sources of these uncertainties are the active volume and
dead layer thicknesses of the detectors (5\,\%), the energy resolution and
energy shift after cross talk correction (3\,\%), the MC simulations (4\,\%)
and the uncertainty on the isotopic abundance (2.5\,\%), where the numbers in
parentheses give the resulting uncertainty on the signal efficiency.  The
combined relative systematic uncertainty on the efficiency is 7.5\,\% for all
decay modes. The systematic uncertainty was investigated for the $0^+_{\rm
  g.s.} - 0^+_1$ transition and assumed to be similar for all decay modes.

A sensitivity study was performed using a toy MC under the assumption of no
signal.  The inverse half-life limit was calculated $10^4$ times with randomly
changing input parameters; $n_k^{\rm rand}$ and $m_k^{\rm rand}$ were each
randomized according to a Poisson distribution.  The expectation values of
this distribution were taken from the background model prediction N$_{\rm MC}$
and $4\, \rm{N}_{\rm MC}$, respectively (Table~\ref{tab:SidebandSummary}).
$\epsilon_k^{\rm rand}$ is randomized with a Gaussian distribution with a mean
of $\epsilon_k$ and a width of $\sigma_{\epsilon_k}$. The median of the
90\,\% quantile inverse half-life limit distribution is taken as the
sensitivity.

Table~\ref{tab:results} shows the input parameter for the statistical analysis
($n_k$, $m_k$ and $\epsilon_k$) in the first 3 columns.  The next two columns
show the Frequentist 90\,\% C.L. lower value for $T_{1/2}$ and the expected
sensitivity $\widehat{T_{1/2}}$ from the toy MC study.  The last two columns show
the Bayesian 90\,\% credibility lower bound on $T_{1/2}$ and the respective
sensitivity $\widehat{T_{1/2}}$.

\begin{table}[h]
\begin{centering}
\caption{\label{tab:results}
          Summary of results for all decay modes. Shown from left to right are
          the input parameters for the likelihood: $n$ - number of events
          within the ROI, $m$ - number of events with in SB(1-4), $\epsilon$ -
          detection efficiency. The following columns show the Frequentist
          lower half-life limit and the sensitivity which contains the true
          value in 90\,\% of the cases.  The last two columns show the
          Bayesian lower limit for half-life values with more than
          90\,\% probability and the respective sensitivity.
}
\begin{tabular}{lccc|cc|cc}
\toprule
 & & & & \multicolumn{2}{c|}{Frequentist \unit[90]{\% C.L.}} & \multicolumn{2}{c}{Bayesian \unit[90]{\% C.I.}}\\
Decay mode &  $n_k$ & $m_k$ & $\epsilon_k$  & $T_{1/2}$ & $\widehat{T_{1/2}}$  & $T_{1/2}$ & $\widehat{T_{1/2}}$  \\
 &  & & [\%] & [\unit[\baseTsolo{23}]\,{yr}] &  [\unit[\baseTsolo{23}]{\,yr}] & [\unit[\baseTsolo{23}]{\,yr}] &  [\unit[\baseTsolo{23}]{\,yr}]\\
\midrule
 $0^+_{\rm g.s.} - 2^+_1$ & 2 & 10 & 0.389 & $>1.6$ & $>1.3$ & $>1.3$ & $>1.2$\\ 
 $0^+_{\rm g.s.} - 0^+_1$ & 5 & 34 & 0.919 & $>3.7$ & $>1.9$ & $>2.7$ & $>1.8$\\ 
 $0^+_{\rm g.s.} - 2^+_2$ branch~1 & 6 & 29 & 0.594 & $>1.7$ & $>1.2$ & $>1.4$ & $>1.1$\\ 
 $0^+_{\rm g.s.} - 2^+_2$ branch~2 & 0 & 2 & 0.092 & $>0.74$ & $>0.64$ & $>0.49$ & $>0.46$\\ 
 $0^+_{\rm g.s.} - 2^+_2$ combined & - & - & - & $>2.3$ & $>1.4$ & $>1.8$ & $>1.3$\\  
\bottomrule
\end{tabular}
\end{centering}
\end{table}

\paragraph{decay mode \bb{2} $0^+_{\rm g.s.} - 2^+_1$: }

The transition to the excited $2^+_1$ has one \unit[559.1]{\,keV}
de-excitation \gadash{ray} and hence a small coincidence efficiency compared
to the other excited state decay modes.  After all cuts 2 events are observed
in the ROI and 2.5 events are expected from the side bands. No signal is found
and the Frequentist analysis yields a 90\,\% lower value on the half-life:
\thalftwo\ \unit[\baseT{>1.6}{23}]{\,yr}.  The sensitivity as defined above is
\unit[\baseT{1.3}{23}]{\,yr}. The Bayesian analysis yields lower credibility
bounds on the half-life of \thalftwo\ \unit[\baseT{>1.3}{23}]{\,yr} (90\,\%
C.I.) with a sensitivity of \unit[\baseT{1.2}{23}]{\,yr}.

This lower half-life limit for the \bb{2} $0^+_{\rm g.s.} - 2^+_1$ transition
is two orders of magnitude better than previous best limits.  However, the
current theoretical prediction are beyond the experimental reach. The most
recent calculation~\cite{Unlu:2014ix} predicts a half-life longer by 4 orders
of magnitude.  The lowest half-life prediction by the Hartree-Fock-Bogoliubov
approach is still a factor of 6 above the current sensitivity. The largest
half-life prediction by shell model calculations is even 7 orders of magnitude
above the current limit.

\paragraph{decay mode  \bb{2} $0^+_{\rm g.s.} - 0^+_1$: }

The transition to the excited $0^+_1$ state has two de-excitation
\gadash{rays} of \unit[559.1]{\,keV} and \unit[563.2]{\,keV} in the final state
and the largest efficiency of the considered decay modes. It also has the
lowest theoretically predicted half-life and is thus of special interest.
After all cuts, 8.5 events are expected in the ROI and 5 events are
observed. No signal is found; a lower half-life value of
\unit[\baseT{3.7}{23}]{\,yr} (90\,\% C.L.) is set for the Frequentist
analysis and \unit[\baseT{2.7}{23}]{\,yr} (90\,\% C.I.) for the Bayesian
analysis.  The sensitivities are \unit[\baseT{1.9}{23}]{\,yr} and
\unit[\baseT{1.8}{23}]{\,yr}, respectively.

This lower half-life limit is 2.5 orders of magnitude better than previous
results for the \bb{2} $0^+_{\rm g.s.} - 0^+_1$ transition. The new limit is
well within the region of theoretical predictions. Bayes factors are
calculated for testing the hypothesis of each NME model in
Table~\ref{tab:PreviousLimits} by taking the ratio $B=p(H_1)/p(H_0)$ in which
$H_1$ is the NME model hypothesis with $T_{1/2}^{\rm Model}$ and $H_0$ the
hypothesis of only background.  The models in
Refs.~\cite{Dhiman:1994gq,Civitarese:1994fd,Stoica:1996np,Aunola:1996hq} have
$B<$\baseTsolo{-6} and are ruled out.  The QRPA model~\cite{Toivanen:1997gz}
has $B=0.001-0.19$ for T$_{1/2}^{}$ = \baseT{(1.0-3.1)}{23}\,yr, respectively.
Recent calculations with RQRPA~\cite{PrivCommSuhonen}, IBM-2
\cite{PrivCommIachello} and SM~\cite{PrivCommMenendez} predict significantly
longer half-lives.  For RQRPA a range can be constrained: $B=0.005$ for $g_A =
1.00$ (T$_{1/2}^{}$ = \baseT{1.2}{23}\,yr) compared to $B=0.45$ for $g_A=1.26$
(T$_{1/2}^{}$ = \baseT{5.8}{23}\,yr).  The IBM-2 and SM prediction are still
above the current experimental reach.

\paragraph{decay mode  \bb{2} $0^+_{\rm g.s.} - 2^+_2$: }

The transition into the excited $2^+_2$ state has two de-excitation branches:
Branch~1 (64\,\% probability) with two \gadash{ray} emissions of
\unit[559.1]{\,keV} and \unit[657.0]{\,keV} and branch~2 (36\,\% probability)
with a single \gadash{ray} emission of \unit[1216.1]{\,keV}.  Branch~1 shows a
significantly larger efficiency due to the higher branching ratio; on the
other hand, the ROI for branch~2 is at higher energy resulting in lower
background level. 7.25 events are expected and 6 events are observed for
branch~1 compared to 0.5 expected and no observed event in branch~2. No signal
is found in either branch and a combined limit is calculated according to
Eq.~\ref{eq:likelihood} with $k=1,2$: The Frequentist 90\,\% C.L. lower value
is \unit[\baseT{2.3}{23}]{\,yr}.  The Bayesian analysis yields a 90\,\%
credibility lower bound of \unit[\baseT{1.8}{23}]{\,yr}.  The corresponding
half-life sensitivities are \unit[\baseT{1.4}{23}]{\,yr} and
\unit[\baseT{1.3}{23}]{\,yr}, respectively.

For the \bb{2} $2^+_{\rm g.s.} - 2^+_2$ transition the lower half-life limit
was improvement by 2 orders of magnitude compared to previous limits.  The
theoretical predictions for this transition are \unit[$>$\baseTsolo{28}]{\,yr}
and cannot be tested with the current sensitivity.

\section{Conclusions}

An analysis for \bb{2} excited state transitions in $^{76}$Ge with the
\gerda\ Phase~I data set has been performed for the three decay modes
$0^+_{\rm g.s.} - 2^+_1$, $0^+_{\rm g.s.} - 0^+_1$ and $0^+_{\rm g.s.} -
2^+_2$. The analysis is performed without blinding, however the automated
choice of cuts is expected to have reduced bias: All cut parameters are chosen
such that the sensitivity calculated from MC simulations is maximized.  No
signal has been found and new half-life lower limits are set for all decay
modes which are at least two orders of magnitude larger than those reported
previously.  Bayes factors are calculated for the predictions of the $0^+_{\rm
  g.s.} - 0^+_1$ half-life with various nuclear models.  Many old NME
calculations could be ruled out.

 The analysis is based on the assumption that only one decay mode is realized
 at a time. This is valid in the present case for the non-observation of a
 signal. Hence, the analysis is performed on each decay mode completely
 independently. However, it should be noted that the results of the different
 decay modes are not decoupled since they proceed through the same
 levels. Particularly the \unit[559.1]{keV} \gadash{line} of the $0^+_{\rm
   g.s.} - 2^+_1$ transition is part of all decay modes. The Frequentist lower
 half-life limits are larger than the sensitivity in all cases. A statistical
 background downward fluctuation in the \unit[559.1]{\,keV} region, as
 observed, has a similar influences on all limits. For additional information
 refer to a more detailed description of this analysis in
 Ref.~\cite{Lehnert:2015}.
 
In the future Phase~II of the \gerda\ experiment it will be possible to
increase the sensitivity further. The target mass of \nuc{Ge}{\rm enr}
detectors will be increased by a factor of two in form of relatively small
BEGes detectors.  The overall background level is expected to be reduced by an
order of magnitude.


\section{Acknowledgments}

 The \gerda\ experiment is supported financially by
   the German Federal Ministry for Education and Research (BMBF),
   the German Research Foundation (DFG) via the Excellence Cluster Universe,
   the Italian Istituto Nazionale di Fisica Nucleare (INFN),
   the Max Planck Society (MPG),
   the Polish National Science Centre (NCN),
   the Russian Foundation for Basic Research (RFBR), and
   the Swiss National Science Foundation (SNF).
 The institutions acknowledge also internal financial support.
 
We greatly acknowledge and thank Jouni Suhonen, Francesco Iachello and Javier
Men\'{e}ndez for providing nuclear matrix element calculations for the
investigated decay modes.

The \gerda\ collaboration thanks the directors and the staff of the LNGS for
their continuous strong support of the \gerda\ experiment. Furthermore we
acknowledge the use of the CPU farm ATLAS of ZIH at TU Dresden for extensive
Monte Carlo simulations.

\section*{References}

\end{document}